\documentclass[%
 reprint,
 amsmath,amssymb,
 aps,
]{revtex4-1}

\usepackage{graphicx}
\usepackage{dcolumn}
\usepackage{bm}

\begin{document}

\preprint{APS/123-QED}

\title{Phase Transition in Dimer Liquids}

\author{Danh-Tai Hoang}%
 \email{danh-tai.hoang@apctp.org}
 \affiliation {%
 Asia Pacific Center for Theoretical Physics,
POSTECH, San 31, Hyoja-dong, Nam-gu,
Pohang, Gyeongbuk 790-784, Korea.}
\author{H. T. Diep}
\email{diep@u-cergy.fr, corresponding author}
 \affiliation{%
Laboratoire de Physique Th\'eorique et Mod\'elisation,
Universit\'e de Cergy-Pontoise, CNRS, UMR 8089\\
2, Avenue Adolphe Chauvin, 95302 Cergy-Pontoise Cedex, France.\\
 }%





\begin{abstract}
We  study the  phase transition in a system composed of dimers interacting with each other via a nearest-neighbor (NN) exchange $J$ and competing interactions taken from a truncated dipolar coupling.  Each dimer occupies a link between two nearest sites of a simple cubic lattice.  We suppose that dimers are self-avoiding and can have only three  orientations which coincide with the $x$, $y$ or $z$ direction. The interaction $J$ is attractive if the two dimers are parallel with each other at the NN distance, zero otherwise.
The truncated dipolar interaction is characterized by  two parameters: its amplitude $D$ and the cutoff distance $r_c$. Using the steepest-descent method, we determine  the ground-state (GS) configuration as functions of $D$ and $r_c$.
We then use   Monte Carlo simulations to investigate the nature of the low-temperature phase and to determine characteristics of the phase transition from the ordered phase to the disordered phase at high temperatures at a given dimer concentration.    We show that as the temperature increases,  dimers remain in the compact state and the transition from the low-$T$
compact phase to the disordered phase where dimers occupy the whole space is of second order when $D$ is small, but it becomes of first order for large enough $D$, for both polarized and non polarized dimers.  This transition has a resemblance with the unfolding polymer transition. The effect of $r_c$ is discussed.
\begin{description}
\item[PACS numbers:64.70.M- ;
64.70.mf ; 64.70.km ]
\end{description}
\end{abstract}

\pacs{Valid PACS appear here}
\maketitle


\section{Introduction}

The problem of the phase transition in polymers is one of the most important domains not only for fundamental sciences such as statistical physics but also in many interdisciplinary areas such as biophysics and biochemistry.
Since the introduction of the renormalization group\cite{Wilson,Zinn} (RG), the understanding of the nature of the phase transition in many systems  becomes clear.
 With its concepts, we understand that the nature of the phase transition in a system depends on a very few parameters such as the symmetry of the order parameter and the space dimension.  Phase transitions in various systems which are very different in nature can belong to the same universality class.  However, there are complicated systems where the RG has much of difficulties in application.  One can mention frustrated systems where competing interactions cause highly degenerate ground states with a combination of several symmetries \cite{Diep2005}.  Polymers and moving interacting species such as dimers are another category of systems which cannot be easily treated with conventional methods.

We are interested here in the phase transition of a system composed of interacting dimers, a kind of axial molecules moving in space.   At low temperatures these dimers are  in an orientationally ordered phase and at high temperatures they are in a liquid state.  There is thus a resemblance to liquid crystals.
We recall that liquid crystals are  somewhere between solid and liquid states where molecules have some spatial orientations which under some conditions can order themselves into some ordered structures.
There are two kinds of ordering in the liquid state: (i) the smectic phase where axial molecules are ordered in planes but no order between planes (ii) the nematic phase where molecules are oriented in the same direction but there is no spatial ordering between them.  The smectic phase is divided in subcategories according to the molecular orientation with respect to the normal axis of the planes.  Of course, at very high temperatures, molecules are completely orientationally disordered: we have the so-called isotropic phase.
Liquid crystals have been a subject of intensive investigations for more than 50 years due to their numerous applications \cite{deGennes,Chandrasekhar}.  Systems of interacting dimers have also been recently studied \cite{Huse,Krauth,Alet,Alet1,Misguich,Papanikolaou}.

  We study in the present paper a model of dimer liquid by taking into account  an exchange interaction between nearest neighboring (NN) dimers.  In addition, in order to create competing interactions between dimers we introduce a truncated dipolar interaction.    Our purpose is not to investigate the effect of the infinite-range dipolar interaction, but to use the competing interactions generated by the two terms of the dipolar interaction within a cutoff distance instead of putting by hand competing interactions between NN, next NN, third NN,  etc.  The dipolar interaction in spin systems in particular in thin films  has been widely studied (see references in Ref. \onlinecite{Santamaria}).    Our aim is to investigate the nature of the ordering and of the phase transition in a liquid crystal described by our model.  This is motivated by recent experiments on periodic layered structures of ordered phases \cite{Galerne,Mach1,Mach2,Johnson,Hirst,Wang} and on orientational phase transitions in various liquid crystals \cite{Takanishi,Jakli,Tripathi,Cestari,Calucci,Cordoyiannis,Chakraborty}.  Some numerical investigations on orientational order have also been published \cite{Achim,Peroukidis,Armas}.

   Our model is described in detail in section II.   The ground-state (GS) analysis of polarized dimers is shown in section III and results of Monte Carlo (MC) simulations are shown  and discussed in section \ref{result1}. The case of non polarized dimers are shown in section \ref{result2}. Concluding remarks are given in section \ref{conclu}.

\section{Model}\label{model}

We consider a system of dimers, each of which lies on a link between two nearest sites on a simple cubic lattice.  By definition, dimers do not touch each other.  The dimer axis can be in the $x$, $y$ or $z$ direction.  The  Hamiltonian is given by the following 3-state Potts model \cite{Baxter}:

\begin{equation}\label{HL}
{\cal H} = -\sum_{(ij,mn)}J(ij,mn) \delta(\sigma_{ij},\sigma_{mn} )
\end{equation}
where $\sigma_{ij}$ is a variable defined for the link between nearest lattice sites $i$ and $j$.  $\sigma_{ij}$ is equal to 1 if the dimer axis is $x$, 2 if it is $y$, and 3 if it is $z$.
We suppose that the interaction  between two dimers $(ij)$ and $(mn)$ is $J(ij,mn)$ and it is equal to $J$ ($J>0$) if they occupy  two parallel links on a square face of a cubic lattice cell,  zero otherwise.
For a description purpose, we shall adopt the following notation to define a dimer: the dimer on the link $(ij)$ is always written with $i$ being the first end. In the case where the dimer is oriented (or polarized), the dimer is considered as a vector.  In the case of non polarized dimers, each dimer is a non oriented segment.
Periodic boundary conditions (PBC) are applied in all directions.

As said in the Introduction, we want to introduce competing interactions between dimers. One way to do is to take the dipolar interaction between dimers within a cutoff distance $r_c$.  The dipolar term is written as

\begin{eqnarray}
{\cal H}_d&=&D\sum_{(ij,mn)}\{\frac{\mathbf{S}(\sigma_{ij})\cdot \mathbf{S}(\sigma_{mn})}{r_{(ij,mn)}^3}\nonumber \\
&&-3\frac{[\mathbf{S}(\sigma_{ij})\cdot \mathbf r_{(ij,mn)}][\mathbf{S}(\sigma_{mn})\cdot \mathbf r_{(ij,mn)}]}{r_{(ij,mn)}^5}\}
\label{dip}
\end{eqnarray}
where $\mathbf r_{(ij,mn)}$ is the vector  of modulus $r_{(ij,mn)}$  connecting the middle point $A$ of the dimer $(ij)$ and the middle point $B$ of the dimer $(mn)$. One has then:  $\mathbf r_{(ij,mn)}\equiv \mathbf r_B-\mathbf r_A$.   In Eq. (\ref{dip}), $D$ is a positive constant depending on the material, the sum $\sum_{(ij,mn)}$
is limited at pairs of dimers within a cut-off distance $r_c$.   A discussion on the dipolar interaction and the cutoff distance is given in the next section. The dimer state is defined by a quantity  $\mathbf{S}(\sigma_{ij})$ given by

\begin{eqnarray}
\mathbf{S}(\sigma_{ij})&=&(S_1,0,0) \  \  \mbox{if}  \  \   \sigma_{ij}=1\\
\mathbf{S}(\sigma_{ij})&=&(0,S_2,0)  \  \  \mbox{if}  \  \   \sigma_{ij}=2\\
\mathbf{S}(\sigma_{ij})&=&(0,0,S_3)  \  \    \mbox{if}  \  \   \sigma_{ij}=3
\end{eqnarray}
There are two cases:

i) the non polarized dimers:

$\mathbf{S}(\sigma_{ij})$ is defined by one of the following non algebraic components $S_1=a$, $S_2=a$, $S_3=a$, $a$ being the lattice constant.

ii) the polarized dimers:

$\mathbf{S}(\sigma_{ij})$ is defined by one of the three-component vectors given above, with $S_1=\pm a$, $S_2=\pm a$ and $S_3=\pm a$.

In this paper we study both non polarized and polarized dimers.  $a$ will be taken to be 1 in the following.

\section{Polarized Dimers: Ground-State Analysis}

\subsection{Without dipolar interaction ($D=0$)}

Without the dipolar interaction, the interaction between neighboring molecules will give rise to an orientational order of molecular axes at low temperatures.  This situation is similar to a 3-state Potts model in three dimensions (3D), but the difference resides in the fact that dimers are moving from one bond to another while in the Potts model, the particle stays at its lattice site. The dynamics which leads to  excited states is not the same: dimers have self-avoiding constraints.  Note that
 the phase transition of the $q$-state Potts model in 3D is of first order for $q > 2$ \cite{Baxter}.  We will see that the 3-state dimer crystal described above undergoes  a second-order transition in the absence of a dipolar interaction.

It is easy to see that  in the GS of the above Hamiltonian  all dimers are parallel, as shown by the dimer configuration of type 1 in Fig. \ref{GSA}.  Note that  if the lattice size is $L\times L\times L$, the number of links is $3L^3$, but the number of dimers $N_d$ in a lattice of linear size $L$ should be less than or equal to $N_d^{max}=L^3/2$ because each dimer takes two lattice sites and they do not touch each other. The dimer concentration $n_d$ is defined as
\begin{equation}\label{nddefinition}
n_d\equiv \frac{N_d}{N_d^{max}}
\end{equation}
If $N_d=N_d^{max}$, i. e. $n_d=1$, then the dimers occupy the whole lattice as shown in Fig. \ref{GSA}.  Now if we use a number of dimers less than   $N_d^{max}$, for instance $N_d=N_d^{max}/2$ ($n_d=0.5$), then in the GS  the dimers will fully occupy the half of the lattice in which they are parallel, leaving the remaining half empty.   In other words, in the GS dimers occupy  the lattice space in a  most compact manner. This GS is similar to that of a polymer on a lattice where it is ``folded" to get a minimal energy at temperature $T=0$ \cite{deGennes,Chandrasekhar}.   We emphasize that when $N_d=N_d^{max}$,  dimers cannot move, the system is blocked. In order to study excitations of  a dimer system at a finite temperature $T$, it is necessary to take $N_d$ less than $N_d^{max}$.  As we will see later, the phase transition temperature as well as some low-$T$ behaviors depend on $N_d$ in both cases $D=0$ and $D\neq 0$.

 \subsection{With dipolar interaction ($D\neq 0$)}
 At this stage, it is worth to discuss about  the dipolar interaction.  The infinite-range dipolar interaction has been used since a long time ago in theories and in simulations.  Often, for the electric  dipolar interaction, the sum was treated by using the Ewald's sum \cite{Ewald,Allen1987} which consists in writing the sum in two sums: the short-range real-space sum and the long-range sum in reciprocal space.  The necessity to perform the second sum is to choose a central unit and make its translation in the whole space to realize a periodic system.
 The central unit  and its translated images have the same average properties.  With some careful techniques to ensure the charge neutrality of the central unit, the Ewald's method can be used with some success. However, unlike the electric dipolar interaction where there is a sum of $1/r^3$ terms, the magnetic dipolar interaction has two sums with $1/r^3$ terms: the second one depends on the spin orientations as seen in Eq. (\ref{dip}).  In  spin systems, the choice of the central unit is not obvious because  the magnetic ordering is no more periodic in the vicinity of the phase transition: spin clusters of all sizes exist in the critical region.  A choice of a periodic unit can alter the nature of the phase transition because replacing the long-range part by a periodic system is more or less equivalent to  replacing surrounding spins beyond a short distance by a same average. This is indeed  a mean-field approach.  It is well-known that low-level mean-field theories do not correctly take into account  critical fluctuations near the phase transition \cite{Zinn},  the critical behavior cannot be thus correctly determined.

 In the present work, as said in the Introduction, our objective is not to treat the infinite-range dipolar interaction.  Our aim is to introduce some competing interactions between dimers.  One way to do  is to introduce by hand competing interactions between NN, NNN, third NN, ...  and to study their effect.  Another way is to choose competing interactions coming from the dipolar formula within  a cutoff distance.
We have chosen the second way and we looked for the effect of the two competing terms of Eq. (\ref{dip}) and of the cutoff on physical quantities. We are convinced that many aspects found here remain in models with limited-range competing interactions.  Results of NN interactions in models of statistical physics allow us to qualitatively understand experimental systems which are not always of short-range interaction \cite{Zinn}.

We see hereafter that when a dipolar interaction between  dimers is introduced, the GS depends on $D$ and $r_c$. In the uniform dimer configuration such as the configuration of type 1 in Fig. \ref{GSA}, the dipolar energy is zero:  when there is only one kind of dimer orientation,  say  axis $z$, the dipolar energy of  $\sigma_{ij}$ is

\begin{equation}\label{Ei}
E_i=D\sum_{mn} [\frac{1}{r_{ij,mn}^3}-3\frac{(u_{ij,mn}^z)^2}{r_{ij,mn}^3}]
\end{equation}
If we transform the sum in integral,  the sum in the first term gives $4\pi \ln r_c$ (integrating from 1 to $r_c$), while the second term gives -4$\pi \ln r_c$, which cancels the first term. This is valid  for $r_c$ larger than 1.
 The energy of the system comes from the short-range exchange term, Eq. (\ref{HL}).

In order to understand the GS found below as functions of $D$ and $r_c$, let us consider a dimer $\sigma_{ij}$ interacting with its neighbor $\sigma_{mn}$.  If they have the same orientation, say the $x$ axis for instance, the energy calculated with Eq. (\ref{dip}) is

\begin{equation}\label{Ex1}
E_i=D[\frac{1}{r_{ij,mn}^3}-3\frac{x_{ij,mn}^2}{r_{ij,mn}^5}]
\end{equation}
in which the first term is positive and the second one is negative.  However when the first dimer is on $x$ and the second one on $y$ for example, the first term in Eq. (\ref{dip}) is zero, the dipolar energy for the perpendicular dimer pair is given by
\begin{equation}\label{Ex3}
E_i=D[-3\frac{S_1x_{ij,mn}S_2y_{ij,mn}}{r_{ij,mn}^5}]
\end{equation}
where $S_1=\pm 1$ and $S_2=\pm 1$.  This term can be positive or negative, depending on the signs of  $x_{ij,mn}$, $y_{ij,mn}$, $S_1$ and $S_2$ of the dimer pair $(ij,mn)$.  So, in the GS, $S_1$ and $S_2$ take their signs which minimize this term.  The difficulty of determining the GS comes from the choice of each dimer pair so as to have the minimum of the global energy. This cannot be analytically done.
To determine the GS, we therefore use the numerical steepest descent method which works very well in systems with uniformly distributed interactions.  This method is very simple\cite{Ngo2007,Ngo2007a} (i) we generate an initial dimer configuration at random (ii) we calculate the local field created at a given dimer by its neighbors using  (\ref{HL}) and   (\ref{dip}) (iii) we change the dimer axis to minimize its energy (i. e. align the dimer in its local field) (iv) we go to another dimer and repeat until all dimers are visited: we say we make one sweep (v) we  do a large number of sweeps until a good convergence is reached.

\begin{figure}
 \centering
  \includegraphics[width=60mm,angle=0]{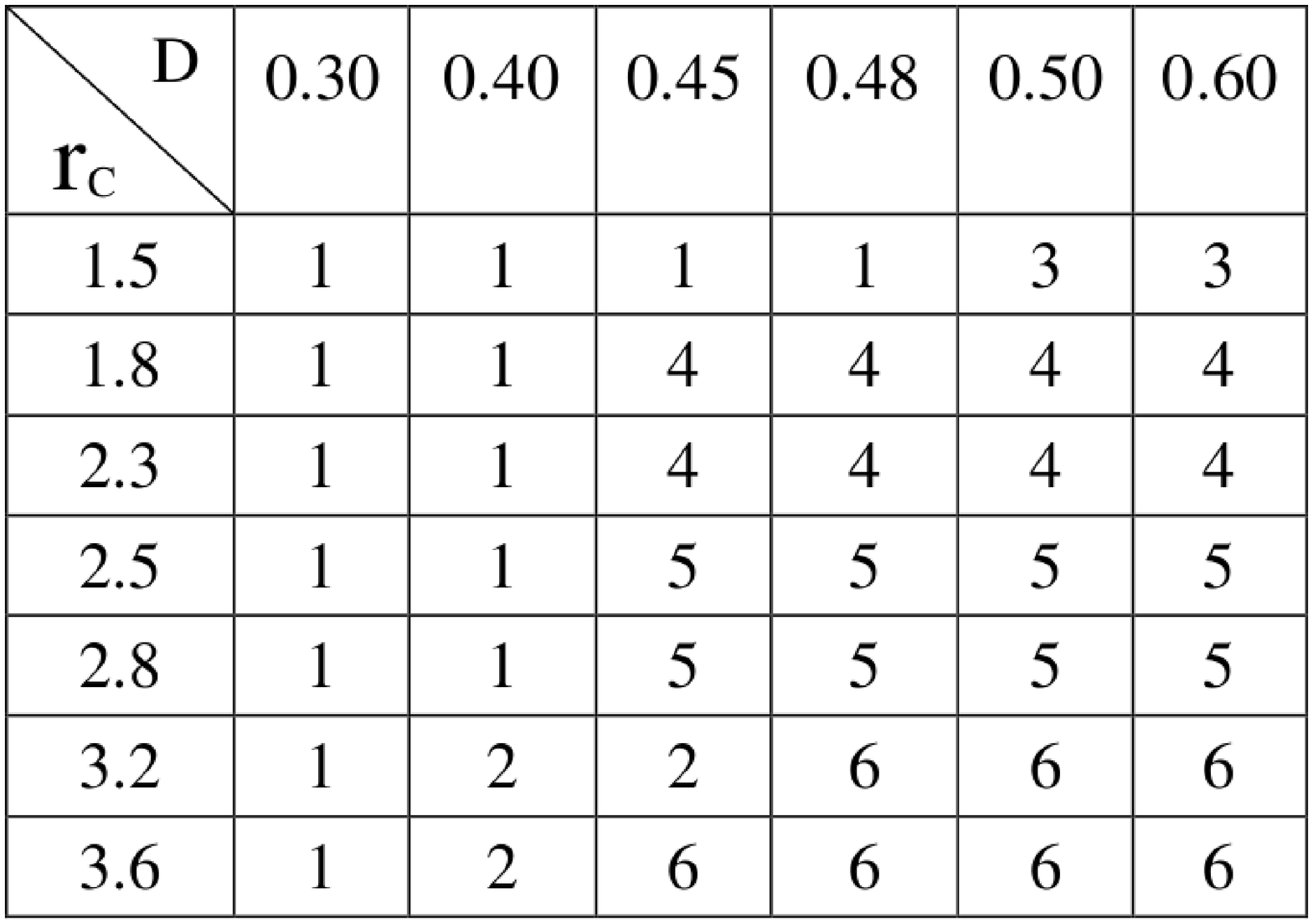}
 \caption{Ground state configurations numbered from 1 to 6 obtained by the steepest-descent method in the space $(D,r_c)$.  These configurations are displayed in Figs. \ref{GSA} and \ref{GSB}.} \label{PD}
\end{figure}

\begin{figure}
\centering
\includegraphics[width=50mm,angle=0]{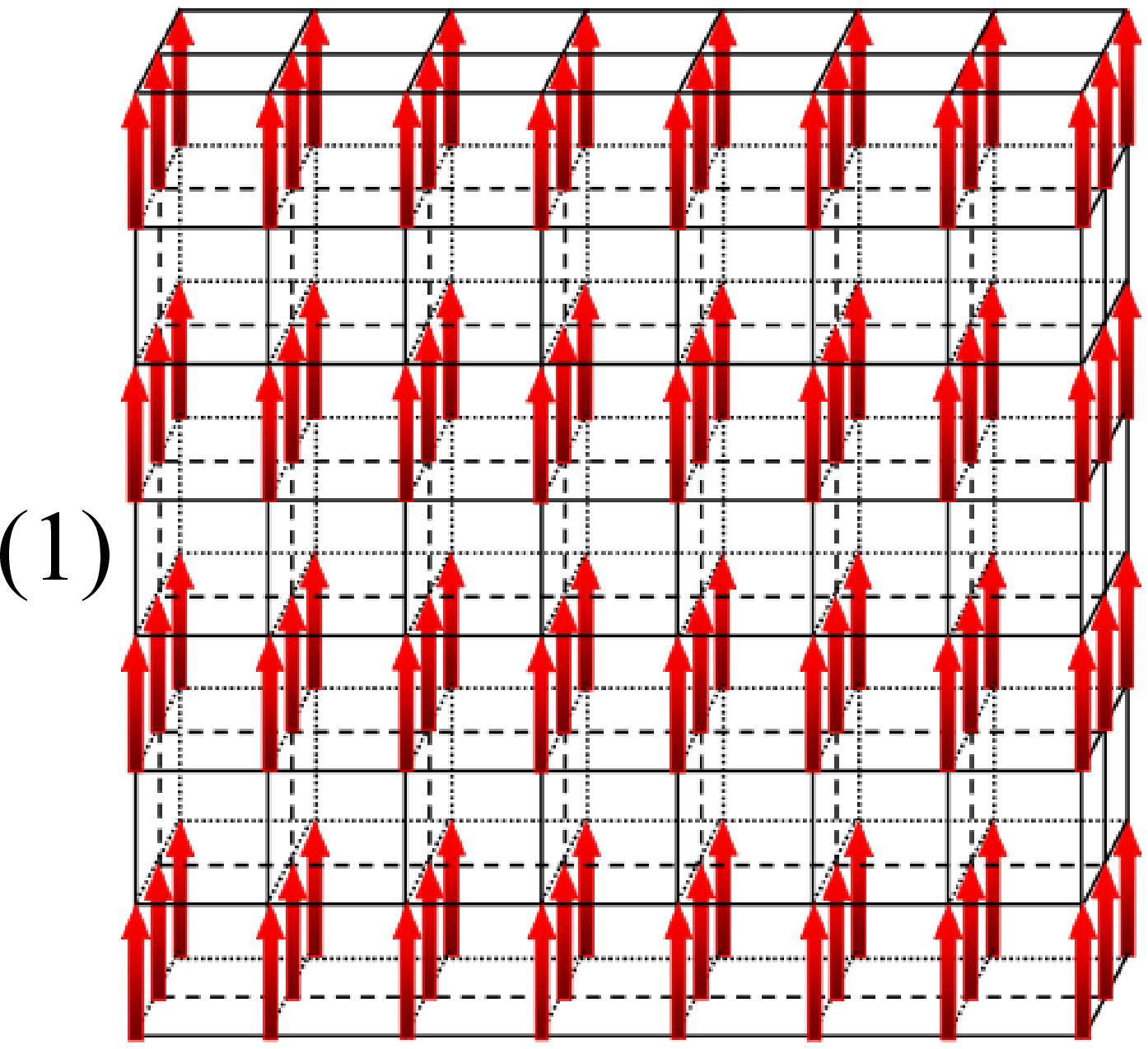}
\includegraphics[width=50mm,angle=0]{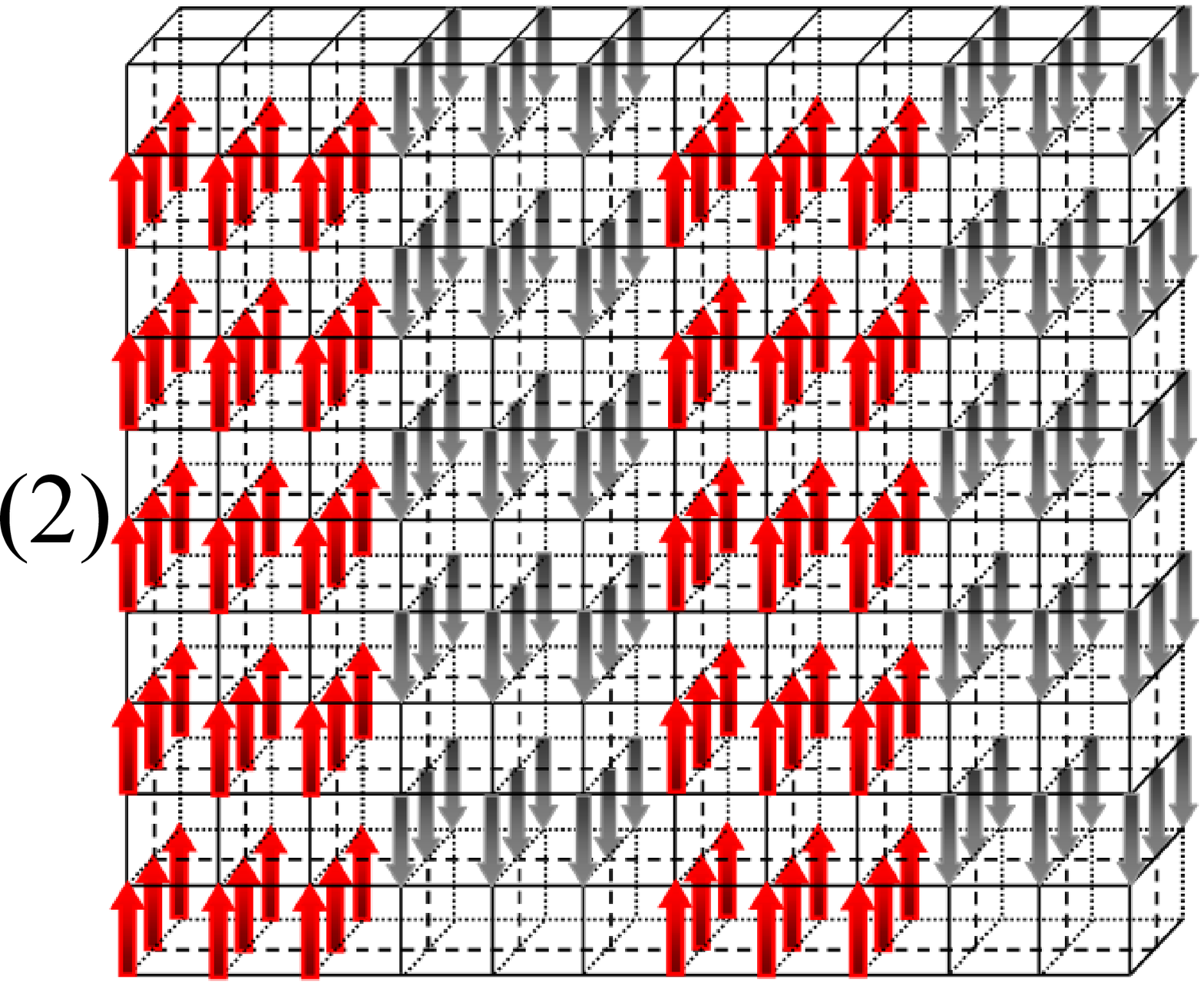}
\includegraphics[width=50mm,angle=0]{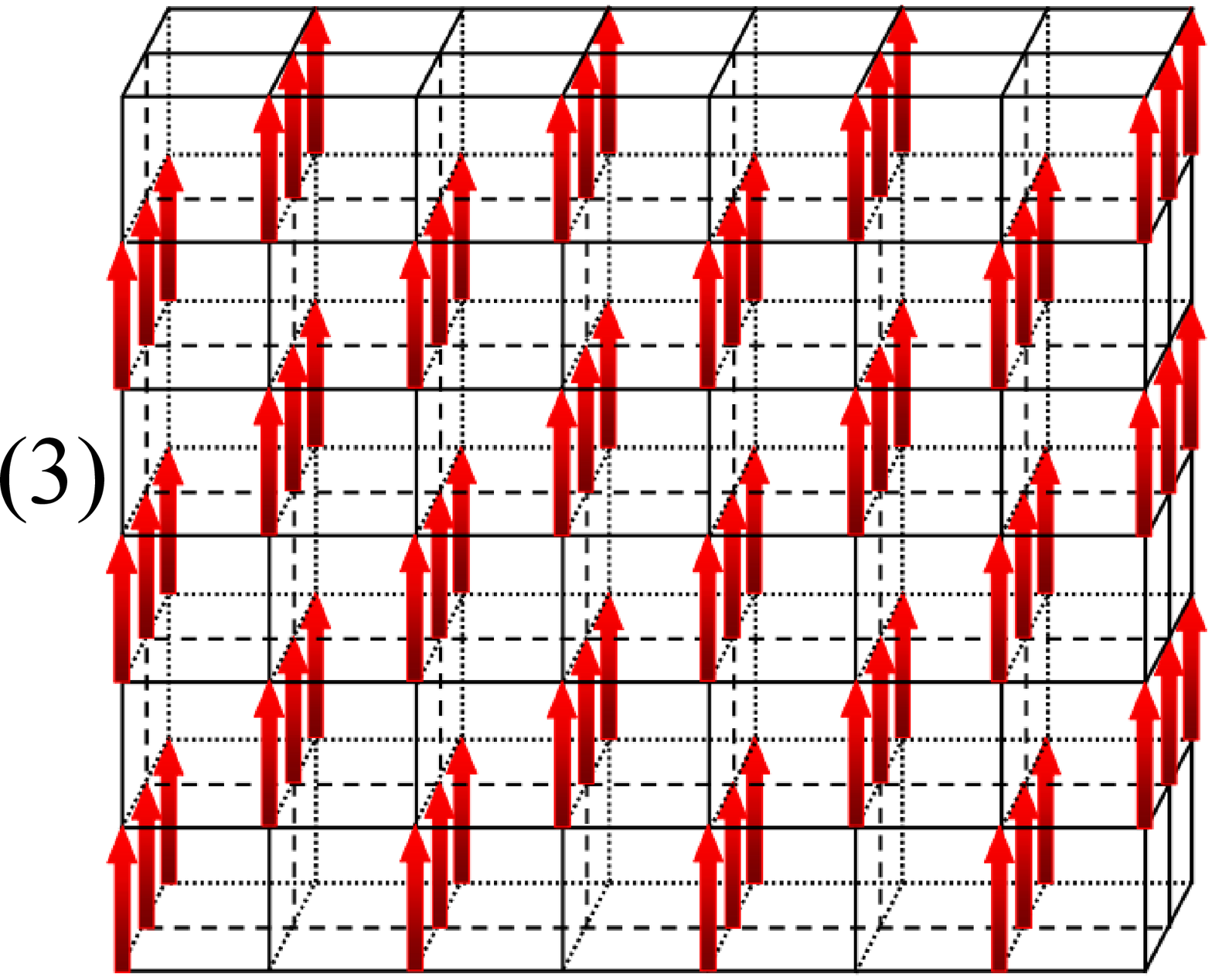}
\caption{(Color online) Ground-state dimer configurations of the type 1, 2 and 3 as indicated in Fig. \ref{PD} in the polarized case are shown.} \label{GSA}
\end{figure}

\begin{figure}
\centering
\includegraphics[width=50mm,angle=0]{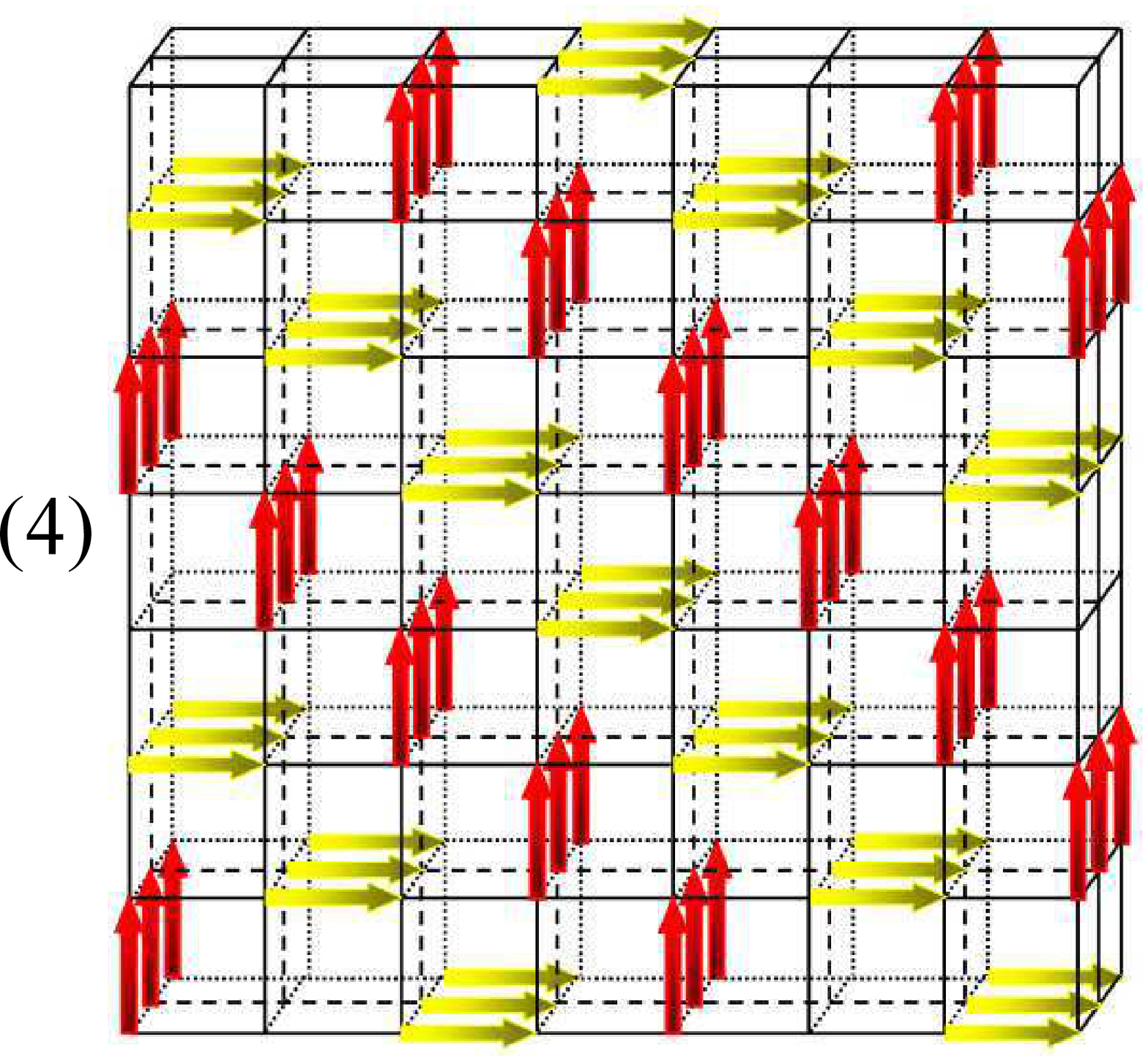}
\includegraphics[width=50mm,angle=0]{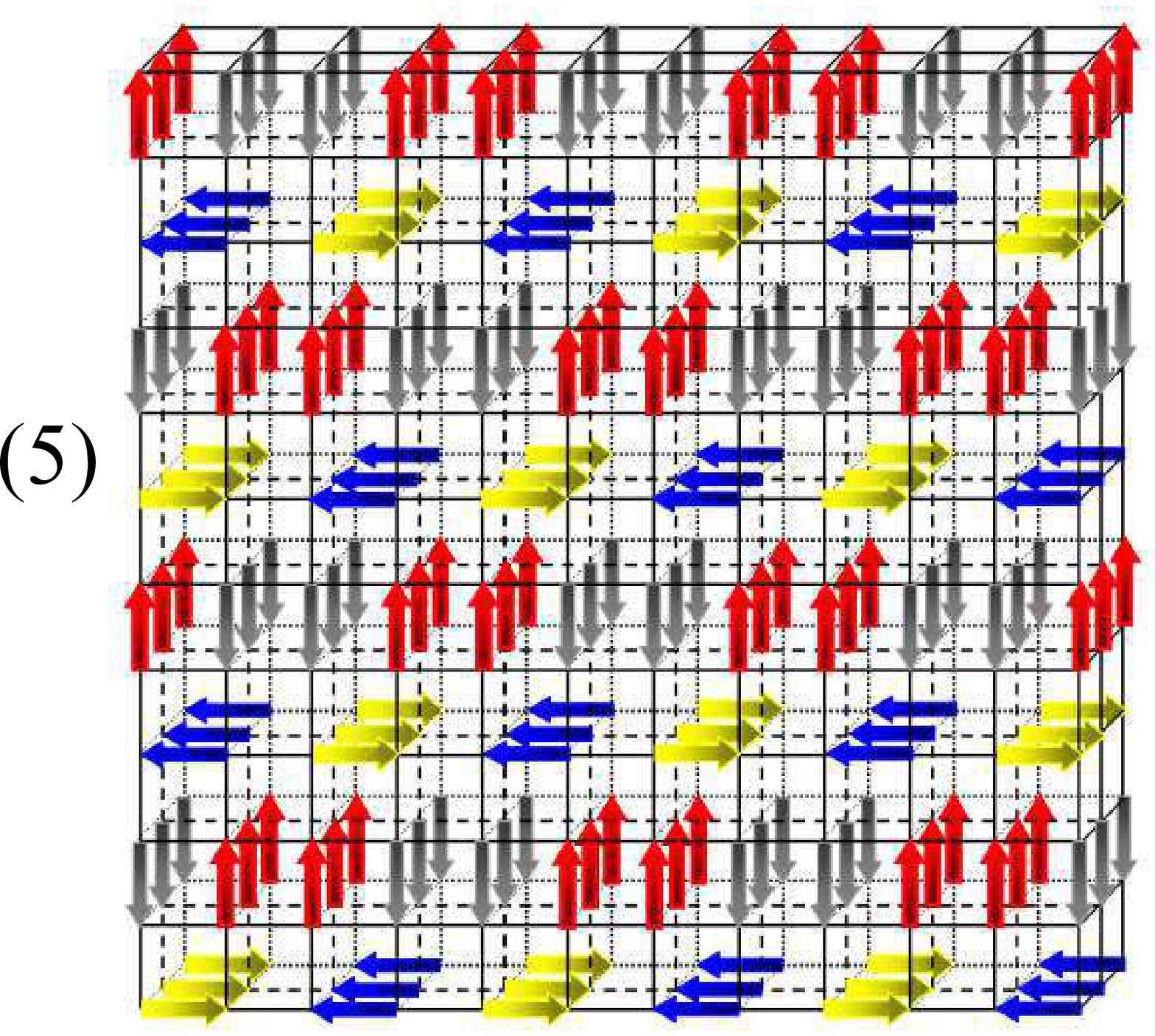}
\includegraphics[width=50mm,angle=0]{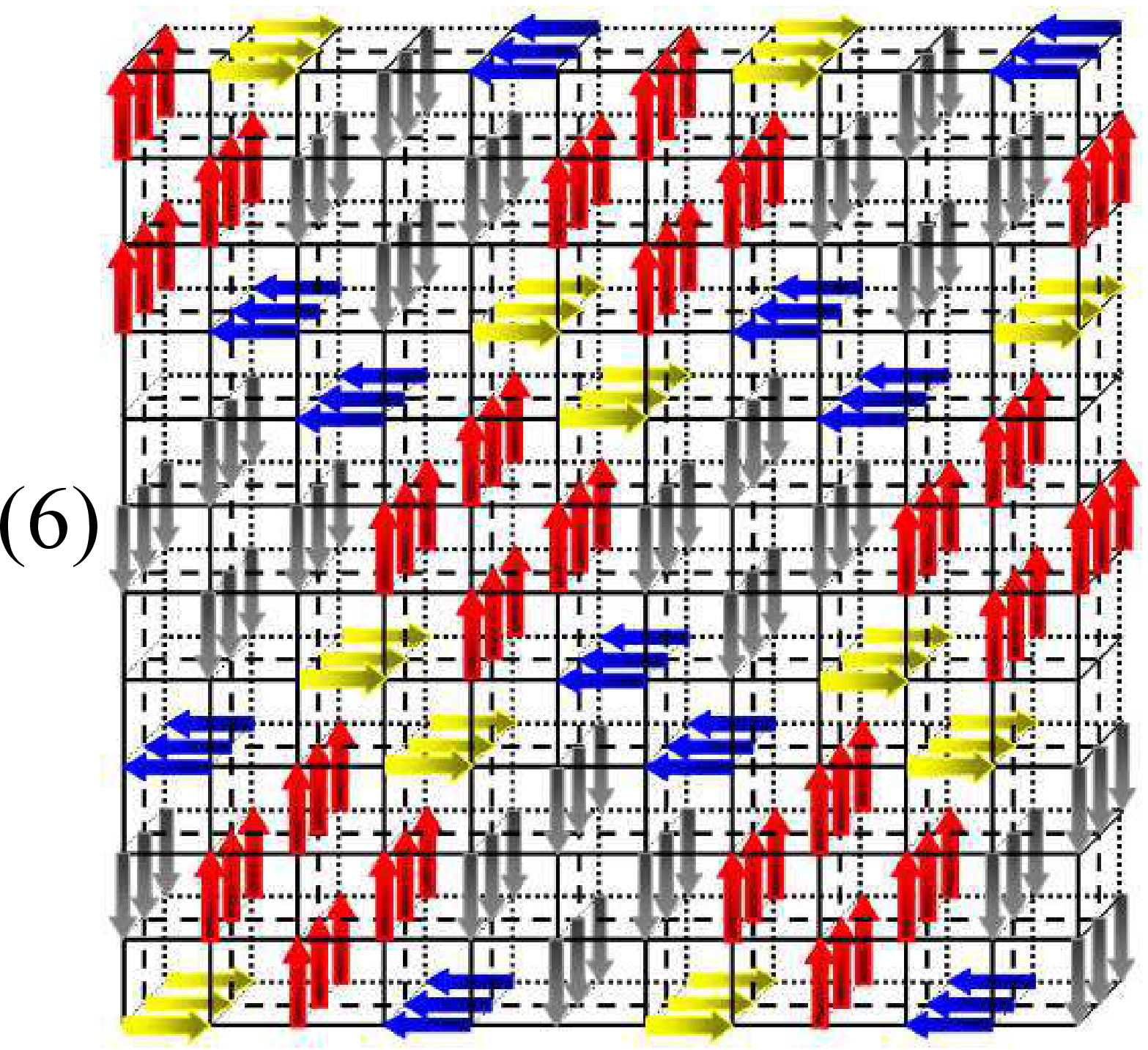}
\caption{(Color online)  Ground-state dimer configurations of type 4, 5 and 6 as indicated in Fig. \ref{PD} in the polarized case are shown.} \label{GSB}
\end{figure}

Using the steepest descent method, we have calculated the GS configurations for various sets of $(D,r_c)$. The results are shown in Fig. \ref{PD}.  For each set $(D,r_c)$, the configuration is indicated by a number.  The configurations corresponding to the numbers from 1 to 6 are shown in Figs. \ref{GSA} and \ref{GSB}.  For a description commodity, let us call $z$ the vertical axis of these figures, and $x$ the horizontal axis.  Note that the GS degeneracy is determined by the number of permutations of the dimer axes.
Let us now comment on Fig. \ref{PD}. For small $D$ and small $r_c$, the GS is of type 1 which is uniform  as in the case $D=0$ discussed above. Larger values of $D$ and $r_c$ yield complicated configurations: for instance, type 2 consists of a three-layered structure of opposite dimer polarizations with a shift in $z$ at their interface , type 3 consists of  a single-layered structure of parallel polarization but with a shift in $z$.  Types 4, 5 and and 6 which are more complicated with larger dimer lattice cells will be presented below.

It is important to note that all the presented GS's are compact where dimers occupy a portion of the lattice. The term ``compact" indicates that there are no free links between dimers. Such GS's are very similar to the folded polymers at low $T$.  The ``volume" of the occupied space with respect to the total lattice volume depends on the dimer concentration $n_d$.  In the figures showing the GS's, only the occupied portion of the lattice is displayed.

\section{Polarized Dimers: Phase Transition}\label{result1}

We consider  a sample size of $L\times L\times L$ where $L$  varies from 12 to 24.  The  NN exchange interaction is used as the unit of energy, i. e. $J=1$.  For the dipolar term, a cutoff distance $r_c$ is taken  up to $\sqrt{10}\simeq 3.15$ lattice distance. At this value, each lattice site has 146 neighboring sites.

The algorithm is described as follows: we consider a dimer and calculate its interaction energy $E_1$ with other dimers within $r_c$ using Eqs. (\ref{HL}) and (\ref{dip}).  We move this dimer to one of the links surrounding its two ends. There are such 10 links in the simple cubic lattice, but not all of them are ``free" at the other end, namely unoccupied by another dimer.  A random free link is chosen, and the new energy $E_2$ is calculated. If $E_2\leq E_1$, the new position of the dimer is accepted, otherwise it is accepted with a probability $\exp [-(E_2-E_1)/(k_BT)]$. This is the well-known Metropolis algorithm \cite{Binder}.  We go next to another dimer and repeat the procedure until all dimers are visited: we say we accomplish one MC step/dimer.   We use several millions of MC steps/dimer for equilibrating and for averaging. The averaged energy and the specific heat are defined by

\begin{eqnarray}
 \langle U\rangle&=&<{\cal H} + {\cal H}_d>\\
C_V&=&\frac{\langle U^2\rangle-\langle U\rangle^2}{k_BT^2}
\end{eqnarray}
where $<...>$ indicates the thermal average taken  at $T$.

We define the order parameter $Q$ by
\begin{equation}\label{Q}
Q=[q\max (Q_1,Q_2,Q_3)-1]/(q-1)
\end{equation}
where $Q_n$ is the spatial average defined by
\begin{equation}\label{Qn}
Q_n=\sum_j \delta (\sigma_{ij}-n)/L^3
\end{equation}
$n(n=1,2,3)$ being the value attributed to denote the axis of the dimer $\sigma_{ij}$ at the link  $(ij)$.
The susceptibility is defined by
\begin{equation}\label{chi}
\chi=\frac{\langle Q^2\rangle-\langle Q\rangle^2}{k_BT}
\end{equation}

For each lattice size we choose the dimer concentration $n_d$ small enough to allow the dimers to move on free links with increasing $T$.  In the following,  we shall use $n_d=5/6\simeq 0.833$ except in subsect. \ref{nceffect} where we study the effect of $n_d$.
Since we work at finite sizes, so for each size we have to determine the ``pseudo" transition which corresponds in general to the maximum of the specific heat or of the susceptibility. The
maxima of these quantities need not to be at the same temperature. Only at the infinite size, they should coincide. The theory of finite-size scaling\cite{Hohenberg,Ferrenberg1,Ferrenberg2} permits to deduce properties of a system at its thermodynamic limit.  We also use the histogram technique to distinguish first- and second-order transitions. The main idea of this technique is to make an energy histogram at a temperature $T_0$ as close as possible to the transition temperature \cite{Ferrenberg1,Ferrenberg2}. Using this histogram in the formulas of statistical physics for canonical distribution, one obtains energy histograms in a range of temperature around $T_0$.   In second-order transitions, these histograms are Gaussian. They allows us to calculate averages of physical quantities as well as critical exponents using the finite-size scaling.  In first-order transitions, the energy histogram shows a double-peak structure.

 \subsection{Without dipolar interaction $D=0$}
\begin{figure}
 \centering
 \includegraphics[width=60mm,angle=0]{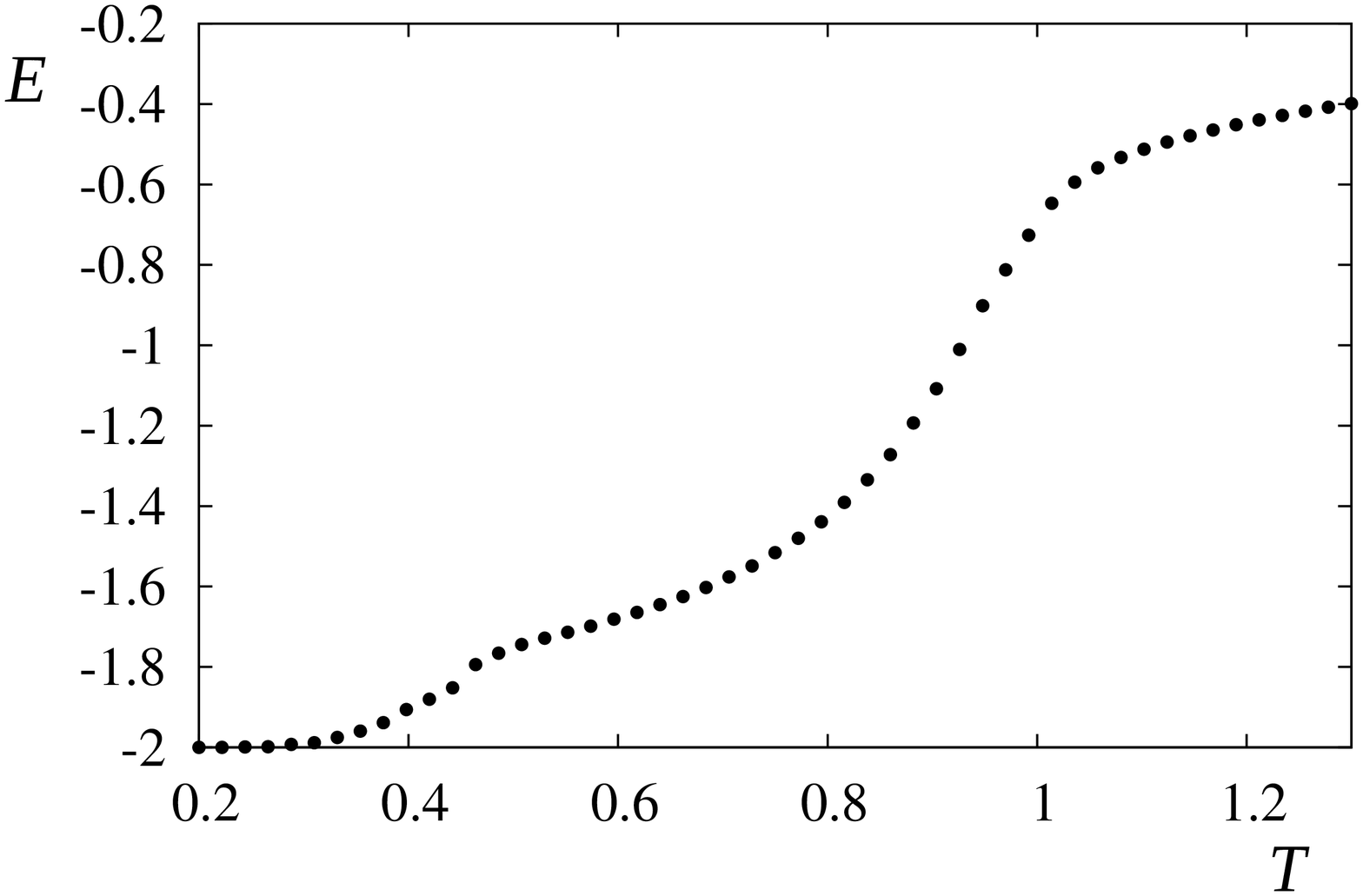}
\includegraphics[width=60mm,angle=0]{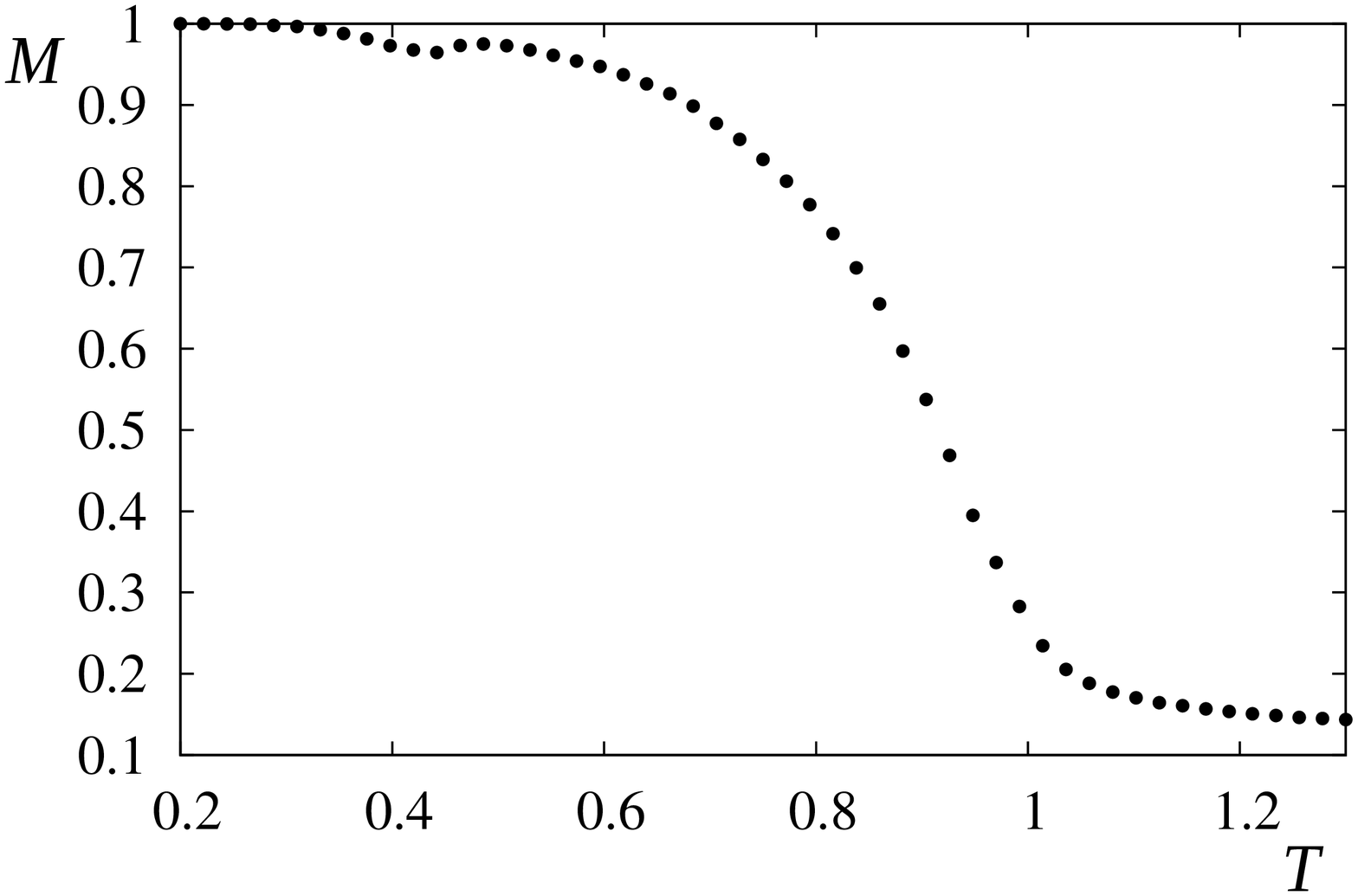}
\includegraphics[width=60mm,angle=0]{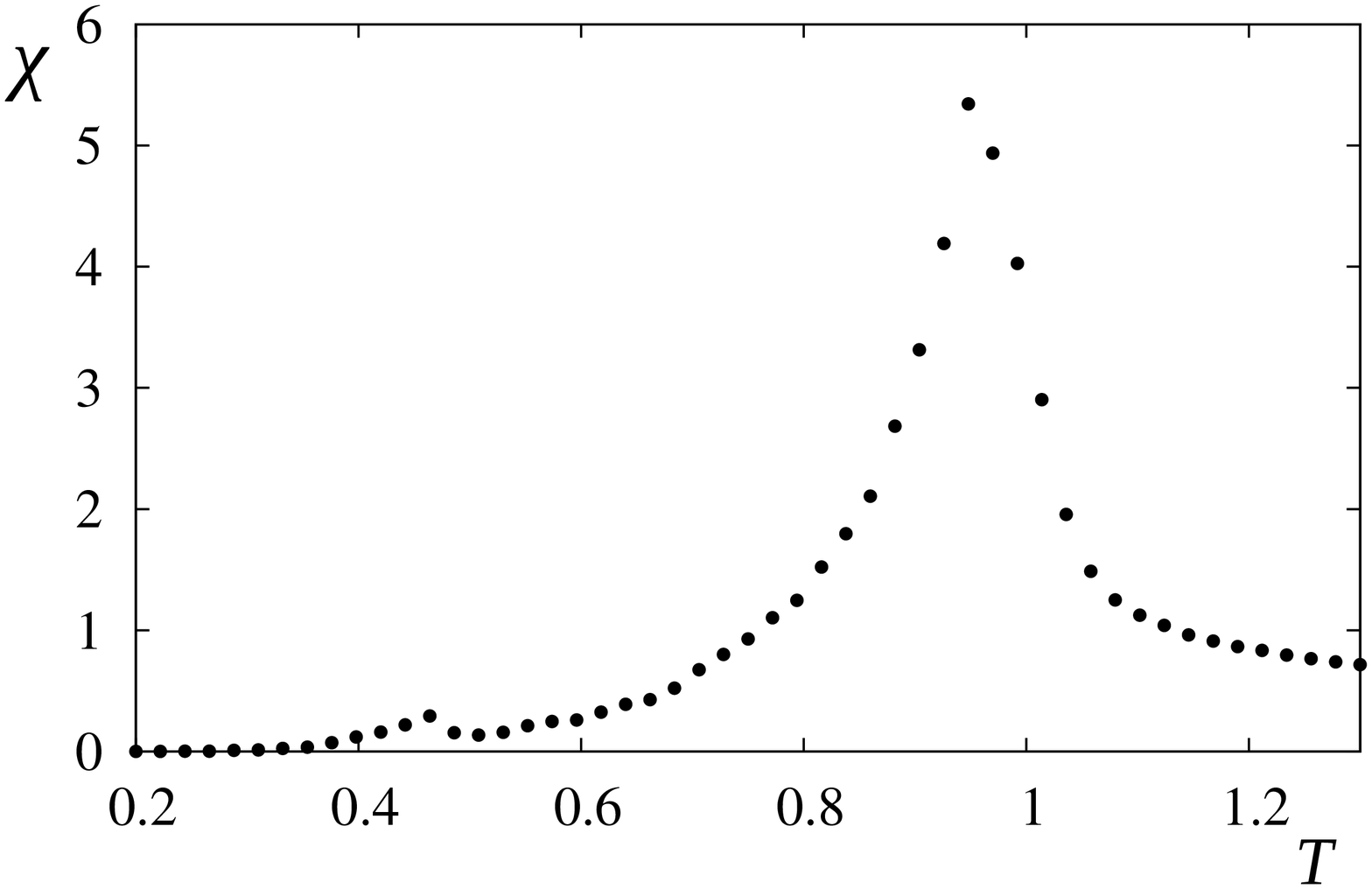}
\includegraphics[width=60mm,angle=0]{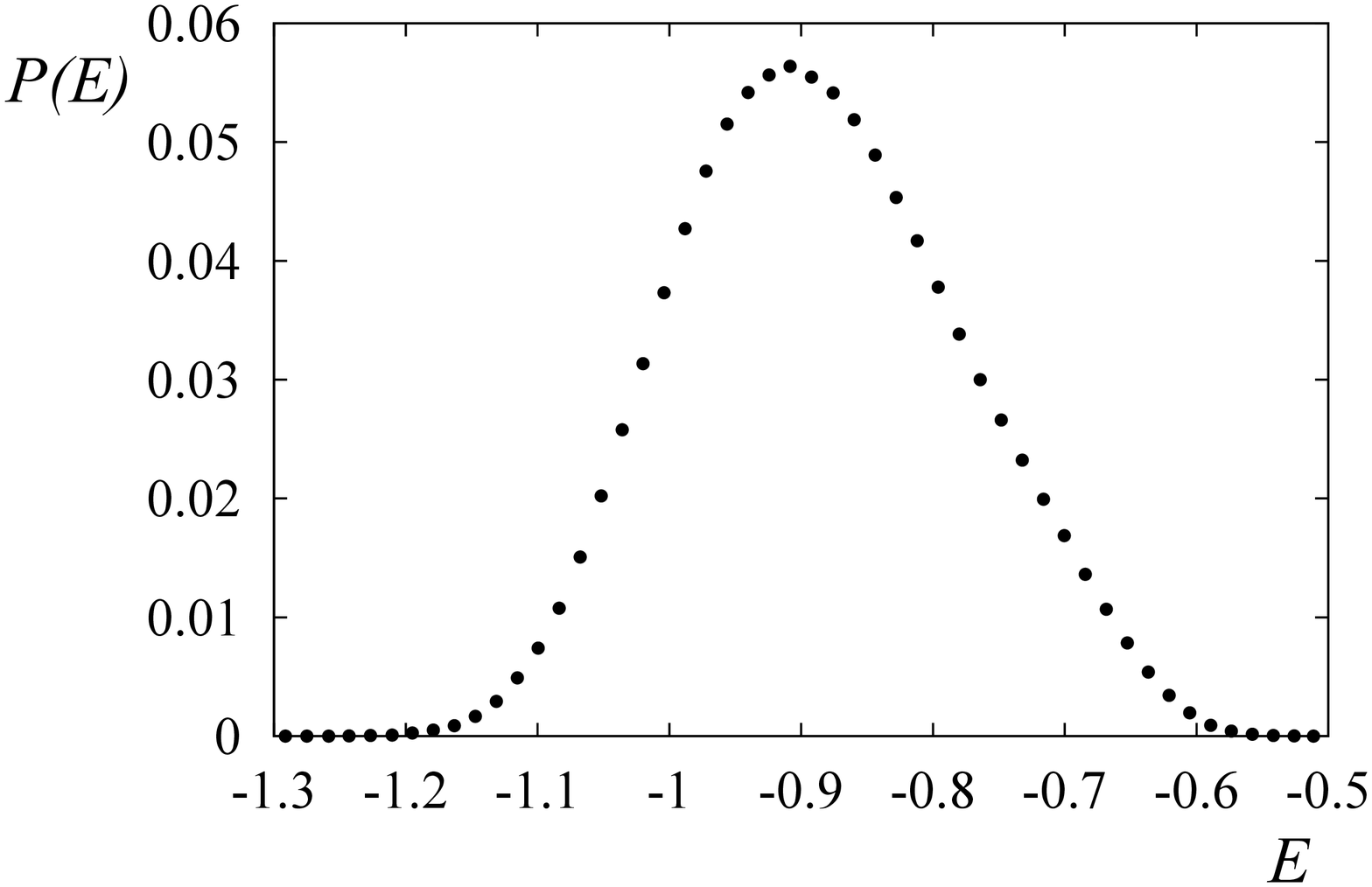}
 \caption{Energy per dimer $E$, order parameter $M=<Q>$ and susceptibility $\chi$ versus $T$ in the case $D=0$. The bottom figure is the energy histogram at the transition temperature.} \label{D0}
\end{figure}

We show first in Fig. \ref{D0}  the energy per site $E\equiv <U>/L^3$,
the order parameter $M=<Q>$ and the energy histogram,  for  $J=1$ and $D=0$.   Several remarks are in order:

(i) At very low $T$ ($<0.4$), the dimers are frozen in the GS configuration.

(ii) For $0.45<T<0.95$ dimers are unfrozen (cf. peak of $\chi$ at $T\simeq 0.45$), they move to occupy free links in the nearby empty lattice space but they remain compact and orientationally ordered.

(iii) For $T>0.95$, they are disordered both in their orientations and spatial positions. The dimers occupy the whole space in the way a polymer does in a solvent at  the unfolding transition.
We find  a second-order transition at $T=T_C\simeq 0.95$ indicated by a Gaussian peak.

\begin{figure}
 \centering
 \includegraphics[width=50mm,angle=0]{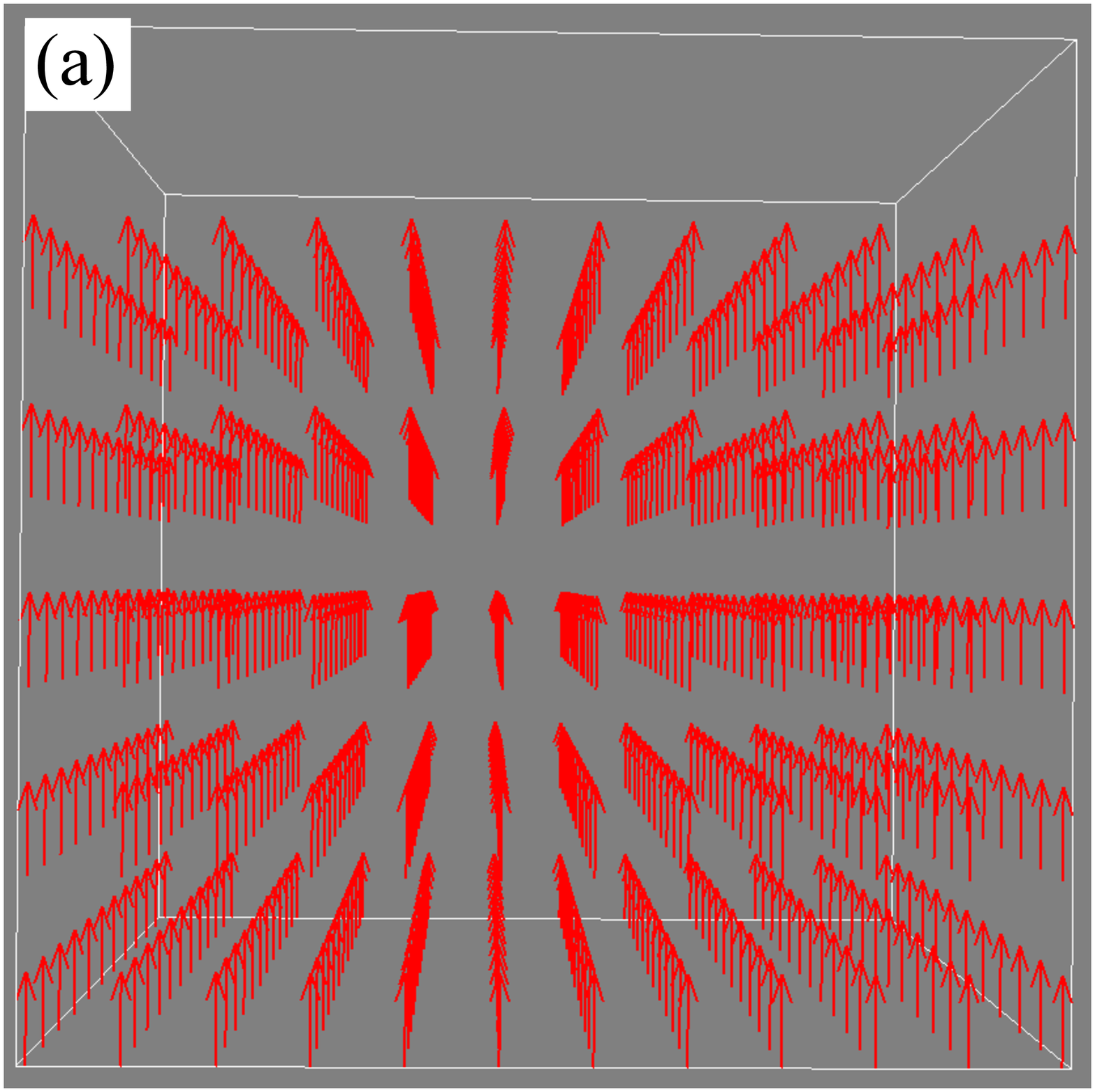}
 \includegraphics[width=50mm,angle=0]{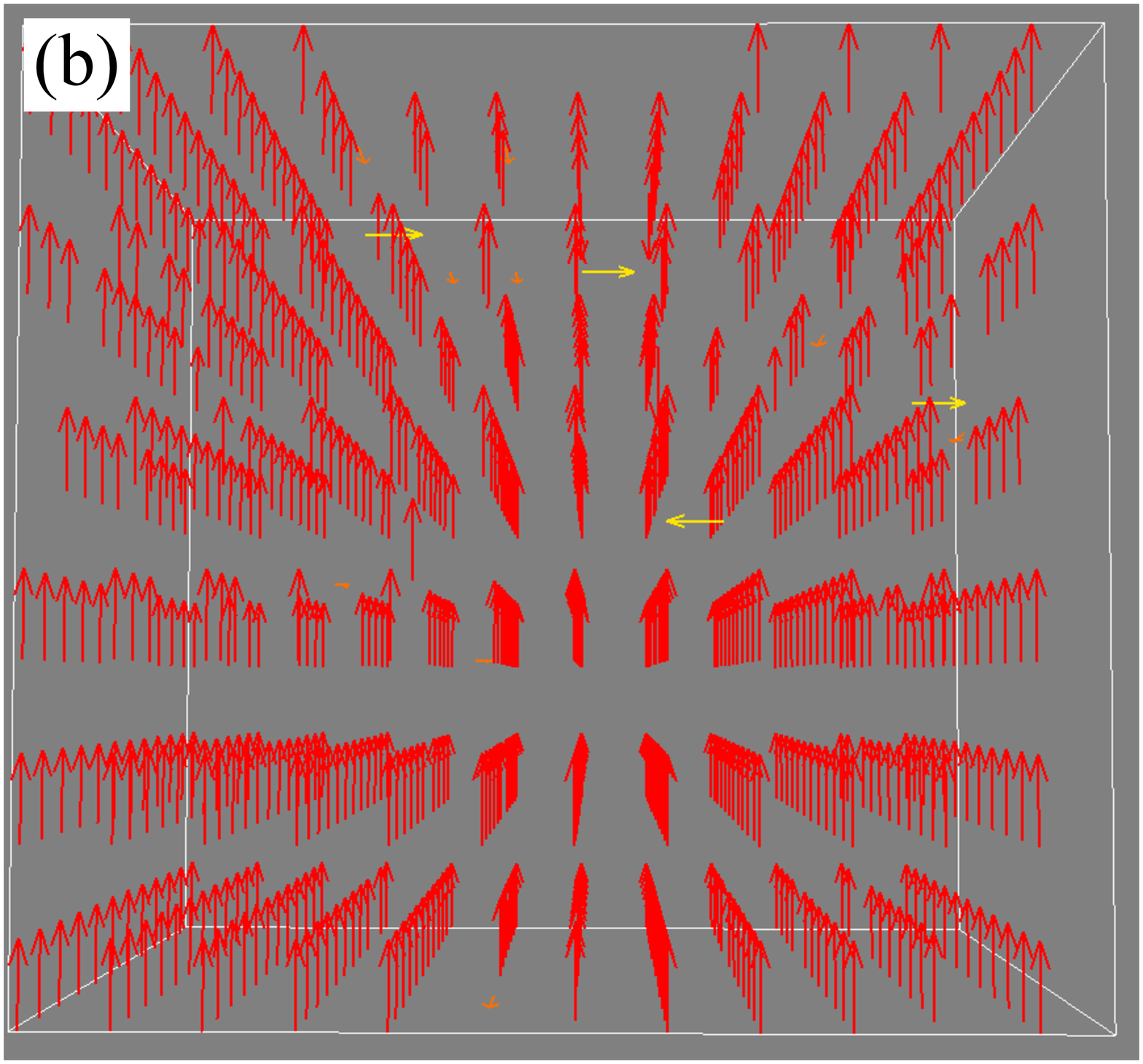}
 \includegraphics[width=50mm,angle=0]{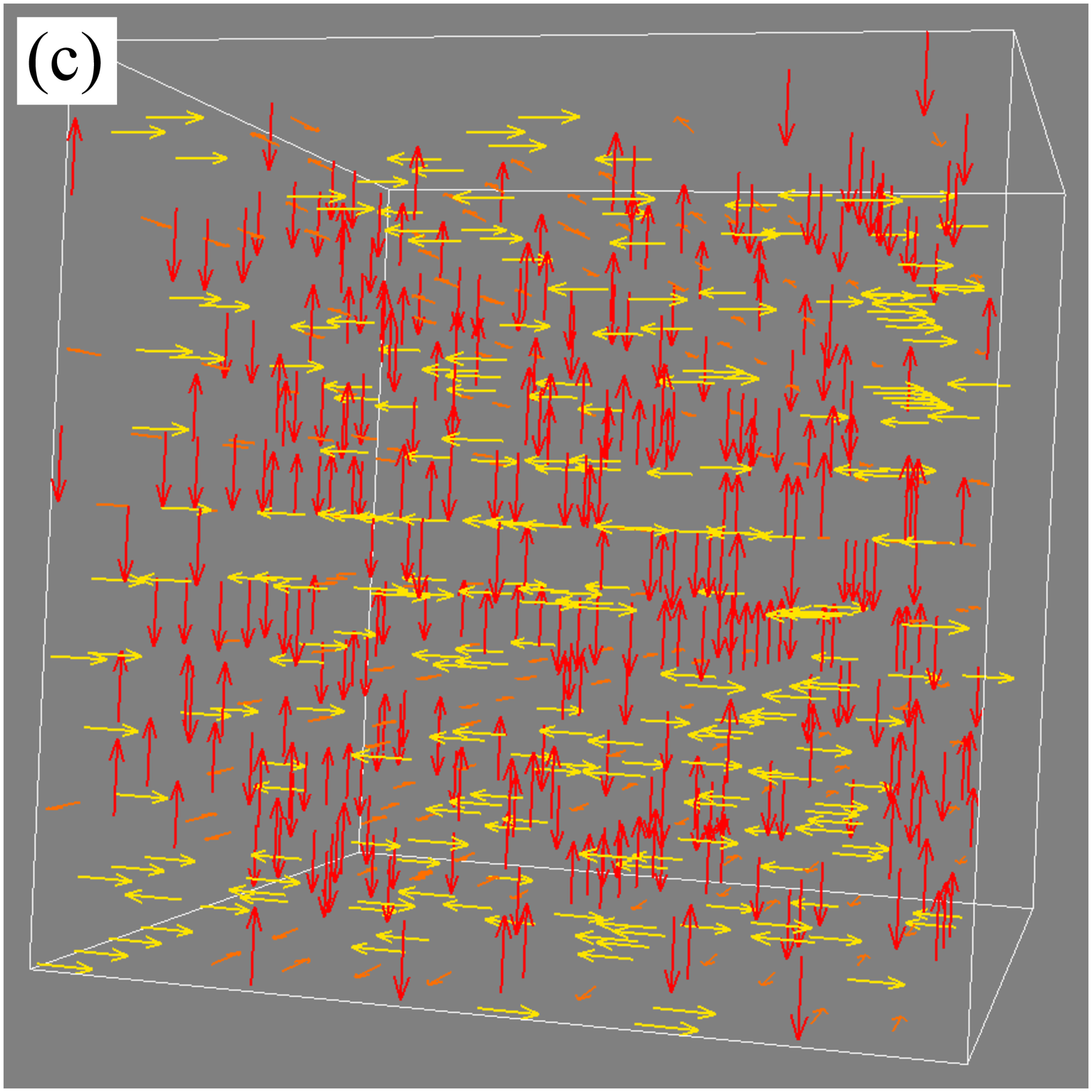}
 \caption{(Color online) Snapshots of the dimer configuration for $D=0$ at (a) $T=0$, ground state (b)  $T=0.508$, compact state (c) $T=1.19$, disordered (isotropic) phase.  See text for comments.} \label{CONFD0}
\end{figure}

We show the snapshots of these phases in Fig. \ref{CONFD0}. In the low-$T$ phase, the upper part
of the lattice is empty (not fully shown) and the lower part is compactly occupied by the dimers.
Only in the disordered phase that the dimers occupy  all the lattice space with vacancies uniformly distributed.

 \subsection{With dipolar interaction $D\neq 0$: polarized case}

\begin{figure}
 \centering
  \includegraphics[width=60mm,angle=0]{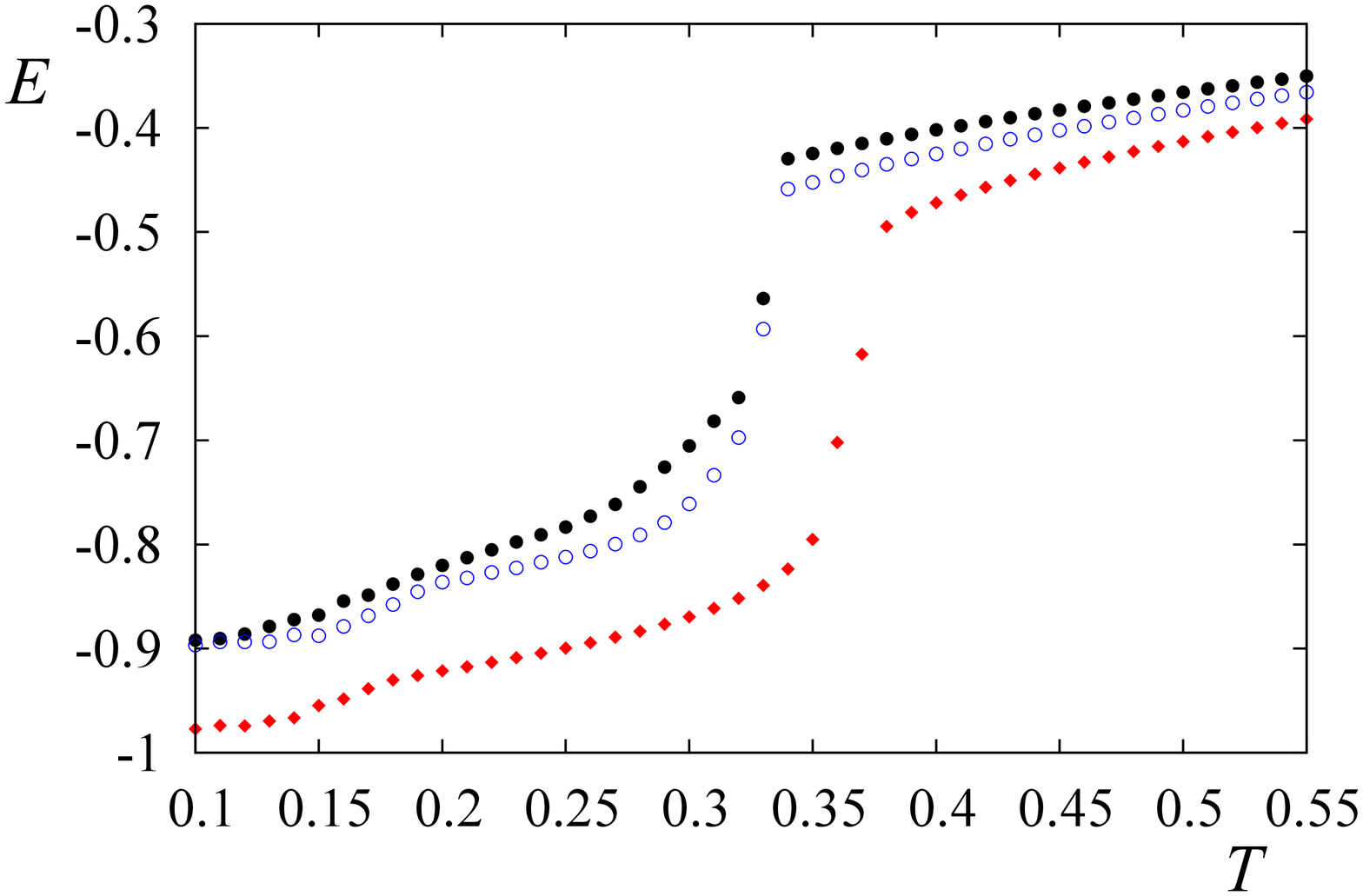}
 \includegraphics[width=60mm,angle=0]{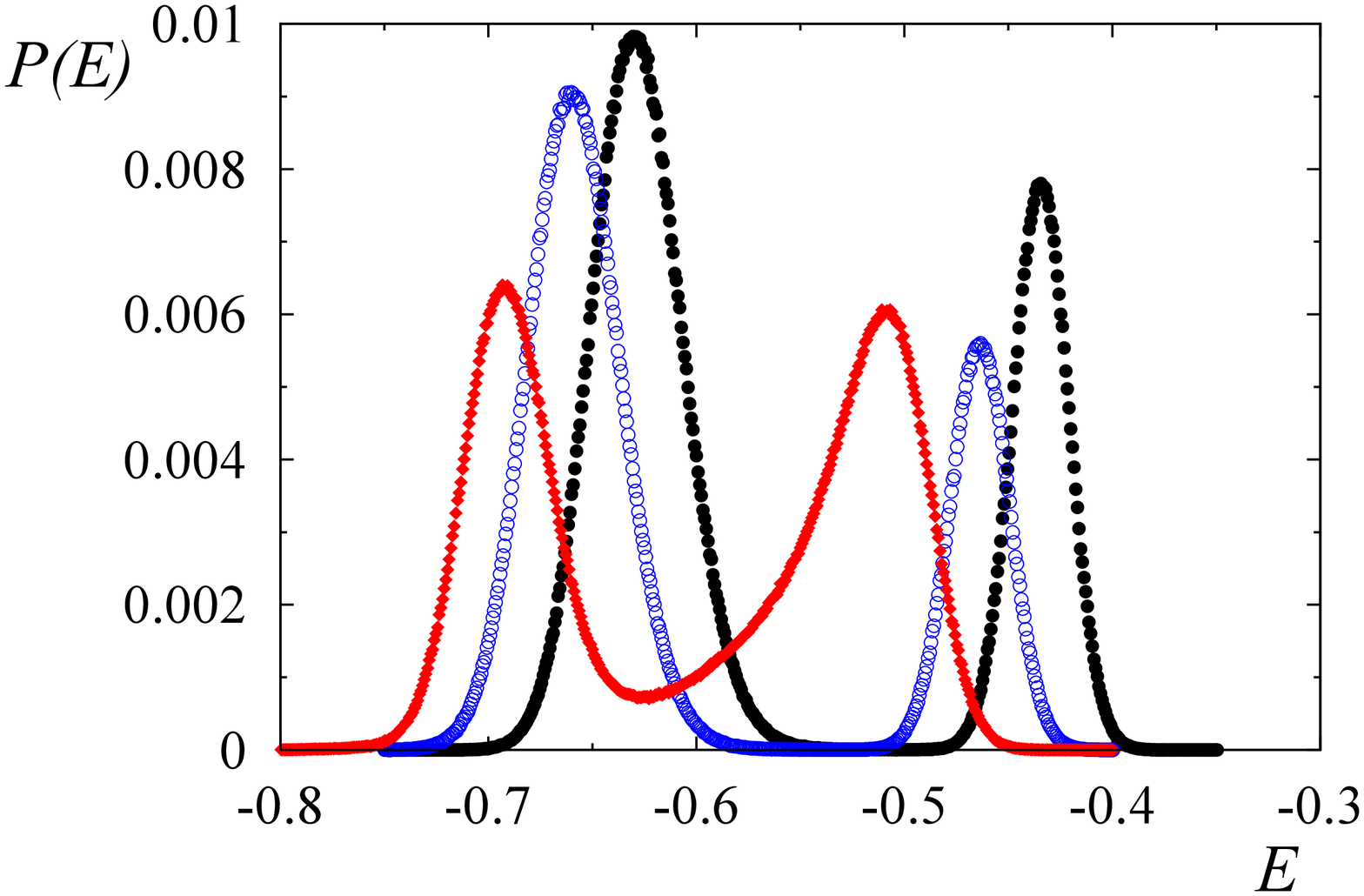}
 \caption{(Color online) Energy versus $T$ and energy histogram at the transition temperature for $D=0.6$ with $r_c=2.3$ (black circles), $r_c=2.5$ (void blue circles) and $r_c=3.2$ (red diamonds).} \label{P}
\end{figure}

The energy versus $T$ for $D=0.6$ and the energy histogram are shown in Fig. \ref{P} for several values of $r_c$ . One observes a discontinuous transition and a double-peak structure which is a signature of a first-order transition, for all $r_c$.

\begin{figure}
 \centering
 \includegraphics[width=50mm,angle=0]{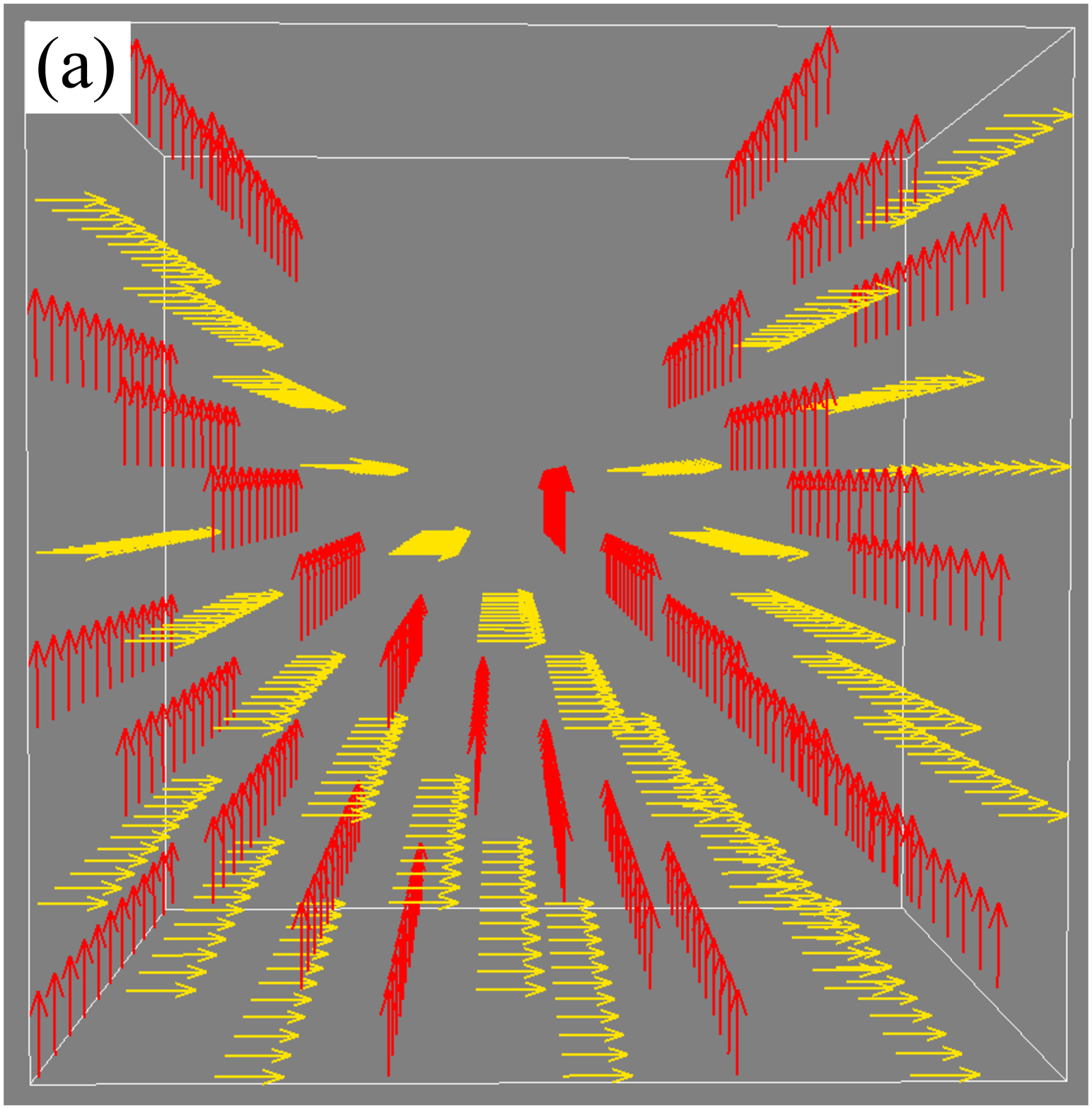}
 \includegraphics[width=50mm,angle=0]{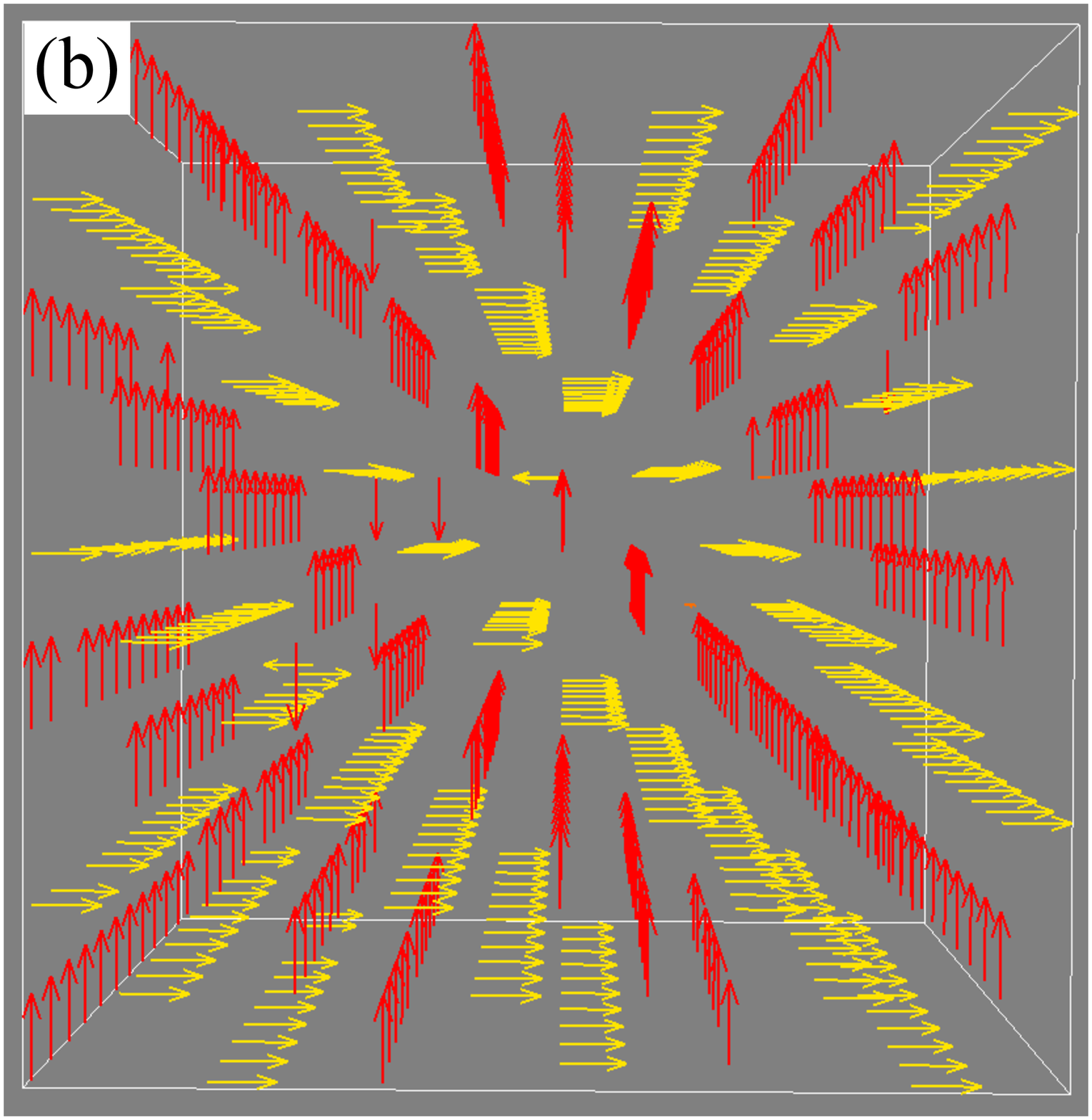}
 \includegraphics[width=50mm,angle=0]{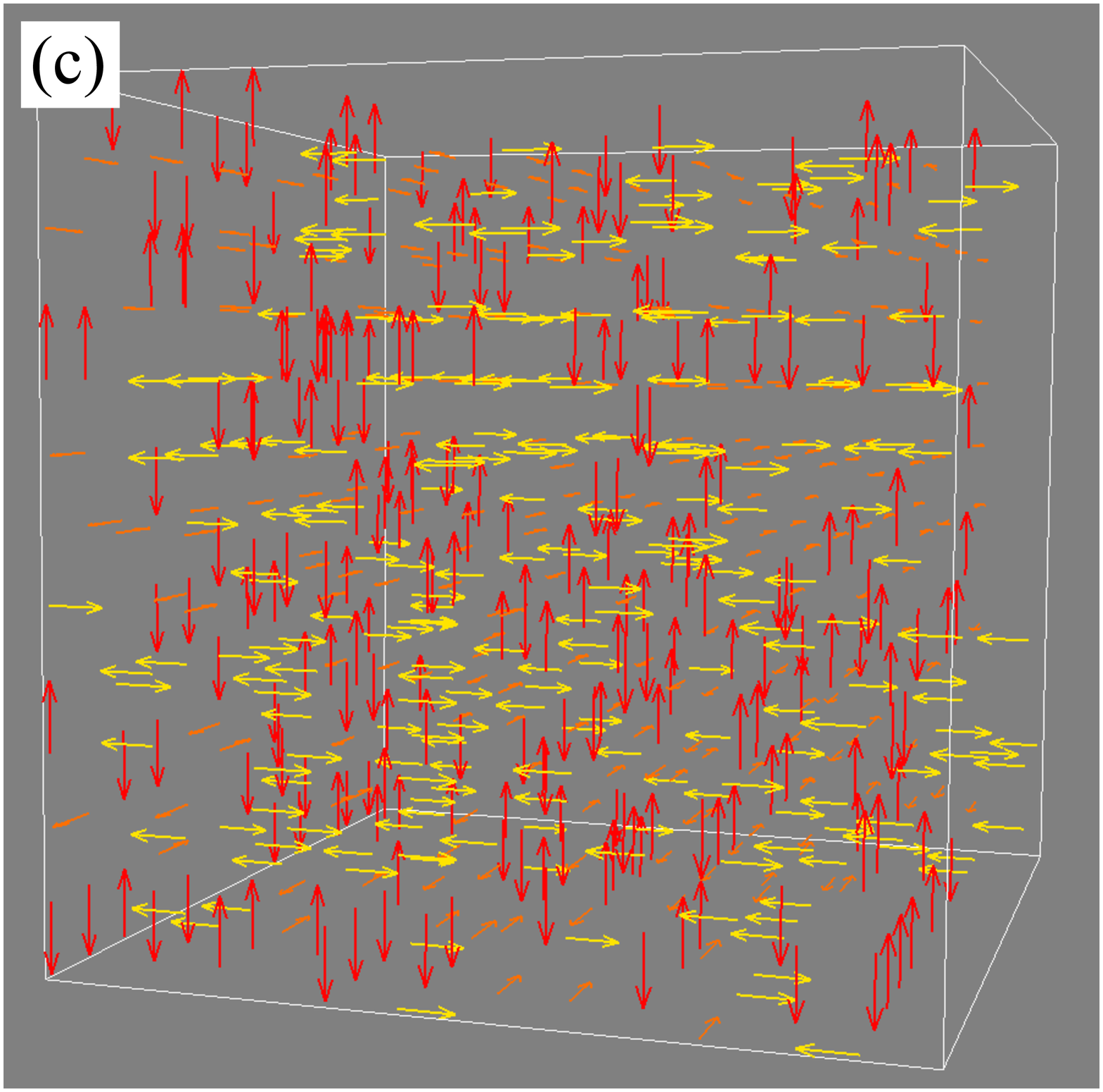}
 \caption{(Color online) Snapshots of the dimer configuration for $D=0.6$ at several temperatures (a) $T=0$, ground state (b) $T=0.17$, compact phase (c) $T=0.45$, disordered (isotropic) phase ($r_c=2.3$).  See text for comments.} \label{CONFD06}
\end{figure}

Let us discuss the case of $r_c=2.3$.  The transition  to the disordered phase
takes place at $T\simeq 0.33$. Note that the system is unfrozen at $T\simeq 0.12$. We show in Fig. \ref{CONFD06} the snapshots of the frozen phase, the compact phase and
the isotropic phase, similarly to the $D=0$ case shown above.   Note that the ``surface" separating occupied and unoccupied parts in the GS is not flat as in the $D=0$ case shown above. In the compact phase, a few dimers move to this  empty space: the system is excited but remains compact.  After the
``unfolding transition",
dimers are uniformly distributed over the whole lattice (Fig. \ref{CONFD06}c).

\subsection{Effect of dimer concentration}\label{nceffect}
\begin{figure}
 \centering
 \includegraphics[width=50mm,angle=0]{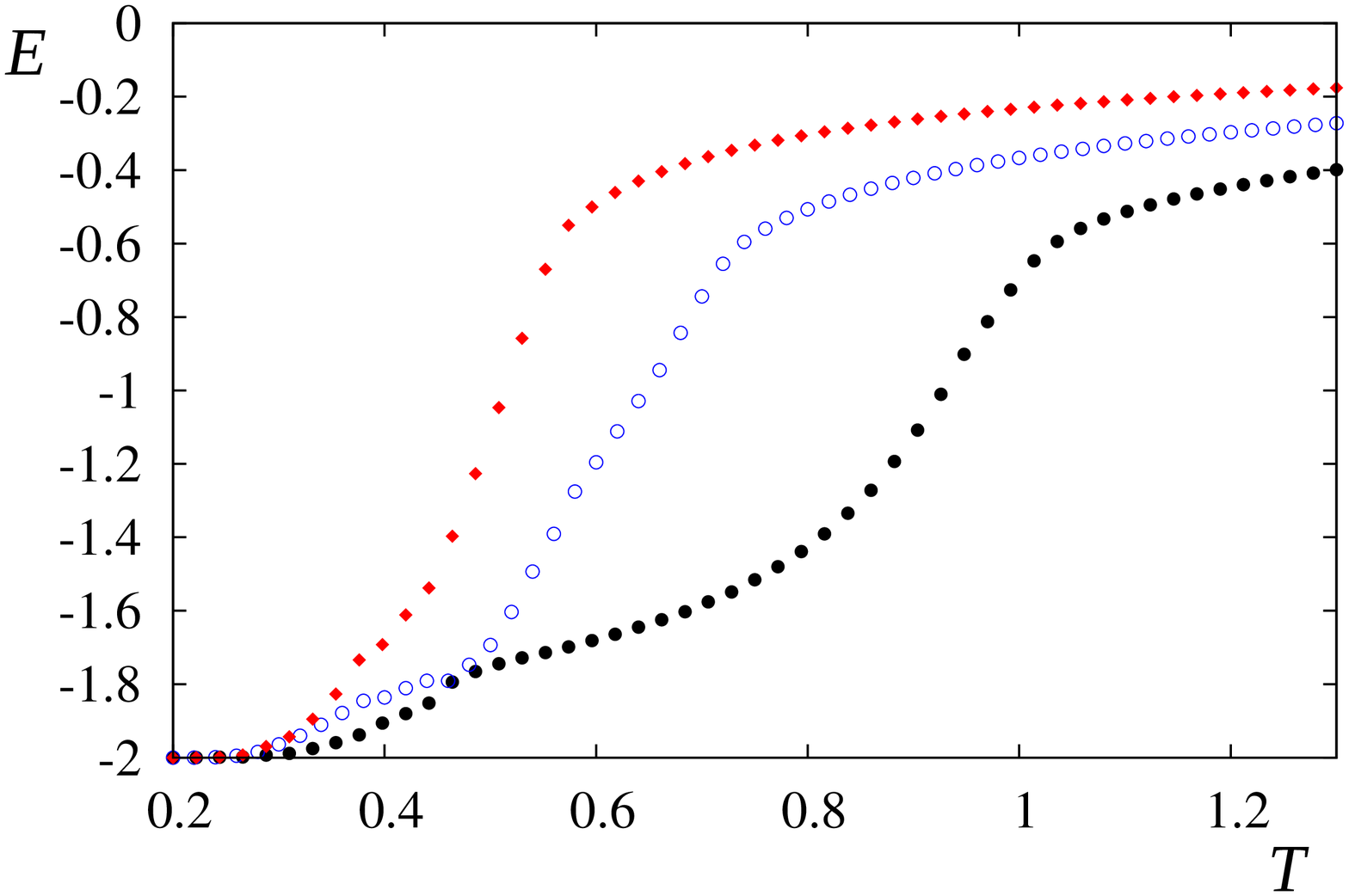}
  \includegraphics[width=50mm,angle=0]{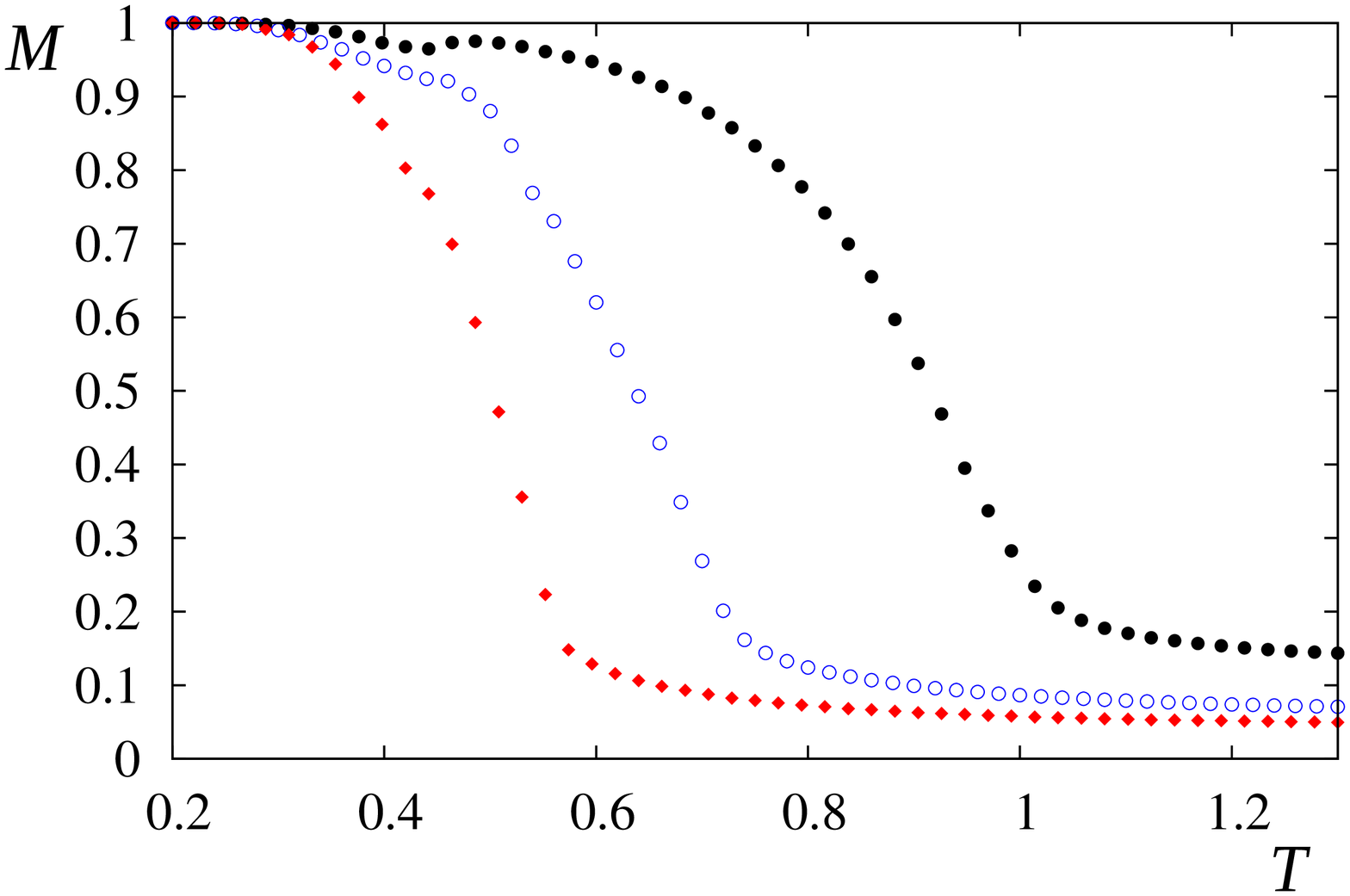}
 \caption{(Color online) Energy per dimer $E$ and order parameter $M=<Q>$ versus $T$ are shown in the case $D=0$  for  concentrations $n_d=5/6\simeq 0.833$ (black circles), $5/8=0.625$ (void blue circles) and $5/12\simeq 0.417$ (red diamonds).} \label{ncD0}
\end{figure}

In this paragraph, we show that the dimer concentration $n_d$ strongly affects the system. The main points are:

(i) Without dipolar interaction: there is a decrease of the transition temperature with decreasing  $n_d$.  This is because the number of neighbors decreases so that the energy per dimer decreases, lowering therefore the disordering temperature.  This is shown in Fig. \ref{ncD0}.

\begin{figure}
 \centering
 \includegraphics[width=50mm,angle=0]{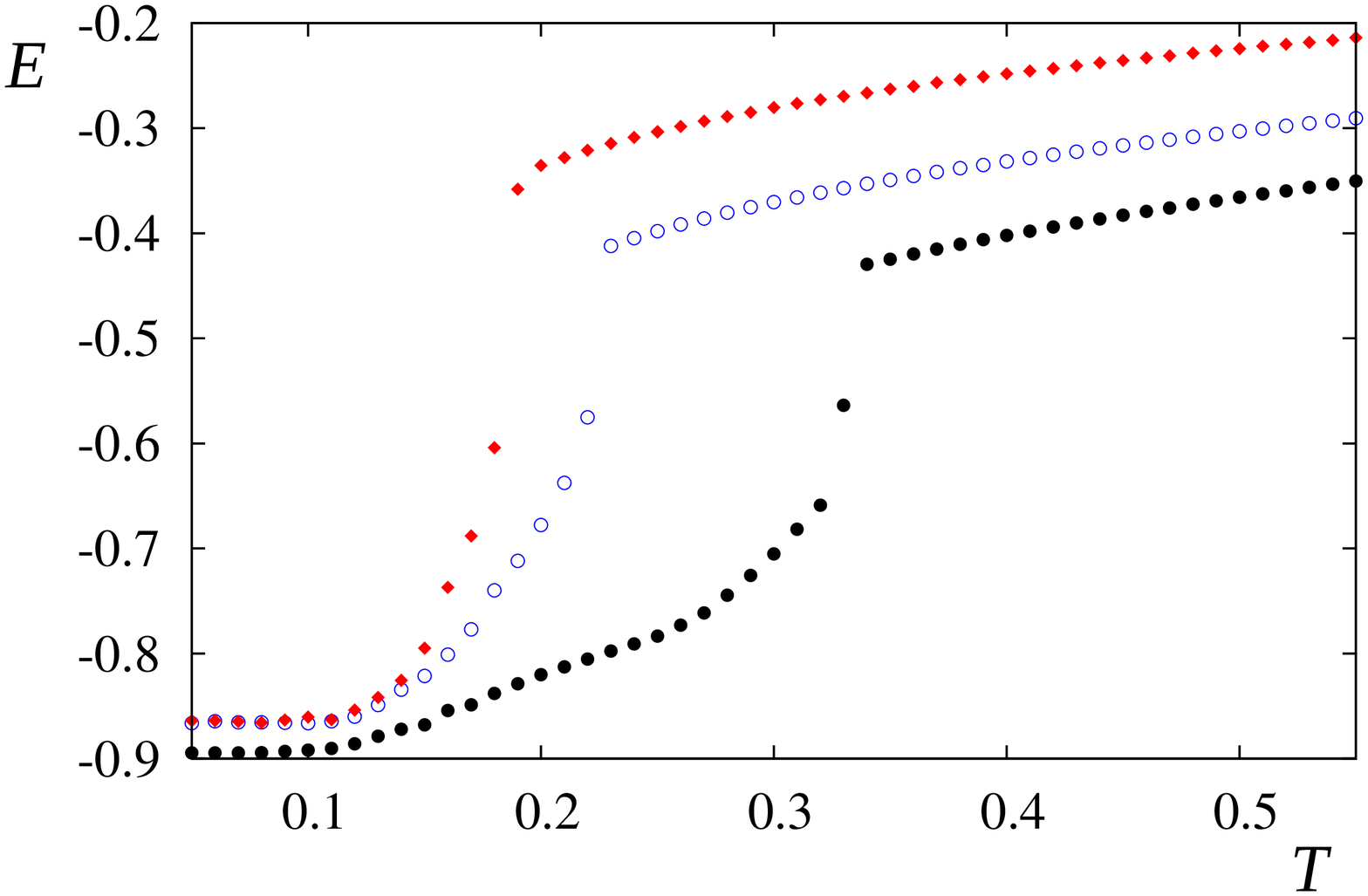}
  \includegraphics[width=50mm,angle=0]{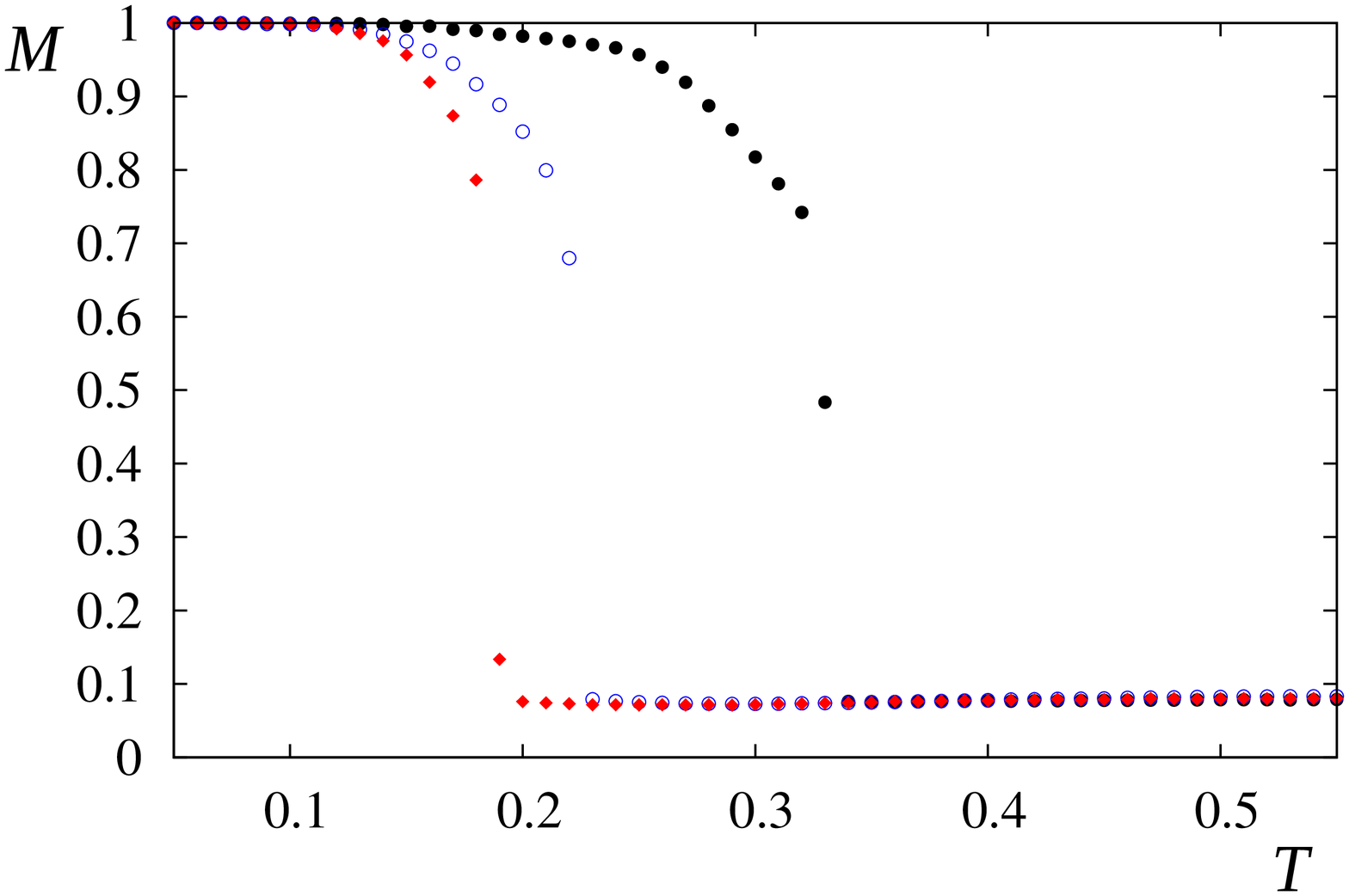}
 \caption{(Color online) Energy per dimer $E$ and order parameter $M=<Q>$ versus $T$ are shown in the case $D=0.6$  with  $r_c=2.3$  for concentrations $n_d=5/6\simeq 0.833$ (black circles), $5/8=0.625$ (void blue circles) and $5/12\simeq 0.417$ (red diamonds).} \label{ncD06}
\end{figure}

(ii) With dipolar interaction: there is a decrease of the transition temperature with decreasing  $n_d$ as in the $D=0$ case,  but  the first-order character observed above remains (see Fig. \ref{ncD06}).

(iii) There is a softening of the unfreezing process at low $T$ ($\simeq 0.42$) in the case where $D=0$: the magnetization becomes smoother for smaller $n_d$ as seen in Fig. \ref{ncD0}. This can be qualitatively understood by observing that the more the empty space is available the easier the unfreezing is realized.  The case of $D\neq 0$ is less visible due to the fact that in this case the interface between occupied and empty
spaces in the GS is much larger, it has a paraboloid form as seen above, making the unfreezing process easy even at high $n_d$.

\begin{figure}
 \centering
 \includegraphics[width=60mm,angle=0]{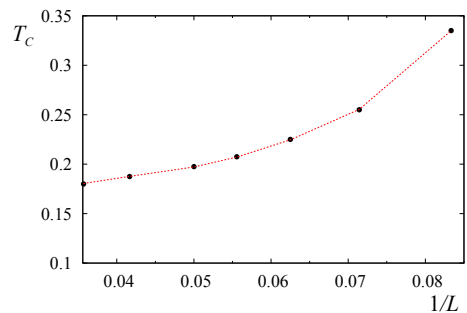}
 \caption{ Transition temperature $T_C$ versus $1/L$ with $L=12$, 14, 16, 18, 20, 24 and 28 ($J=1$, $D=0.6$, $r_c=2.3$).
 See text for comments.} \label{TcNP}
\end{figure}

Note that to decrease
$n_d$ we can increase $L$ while keeping $N_d$ unchanged [see Eq. (\ref{nddefinition})].  We show in Fig. \ref{TcNP} that the transition temperature decreases with increasing $L$ but it remains
finite as $L\rightarrow \infty$.

\section{Non Polarized Dimers: Ground State and Phase Transition}\label{result2}
The main difference of polarized and non polarized dimers resides in the fact that a polarized dimer has an additional internal degree of freedom which makes more abundantly the number of the GS and more excited states at finite $T$. As a consequence, the number of GS in the non polarized case is smaller, the GS configurations are simpler and the transition temperature is higher than in the polarized case.

 \subsection{Ground state}

  In the following we show results for the dimer concentration $n_d=5/6$ similar to main results shown for the polarized case above.

\begin{figure}
 \centering
  \includegraphics[width=60mm,angle=0]{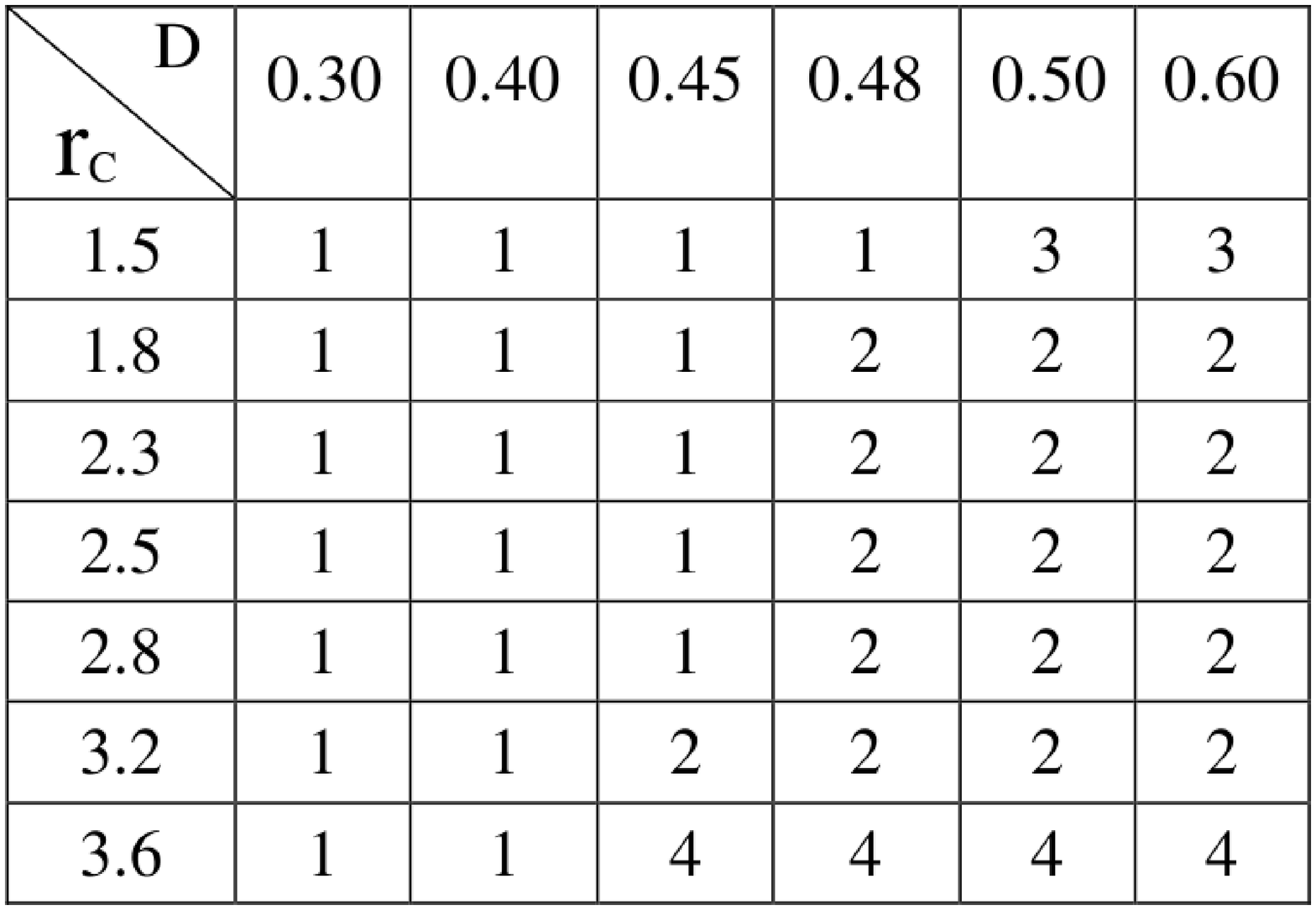}
 \caption{Non polarized dimers: Ground state configurations numbered from 1 to 4 obtained by the steepest-descent method in the space $(D,r_c)$.  These configurations are displayed in Fig. \ref{NPGS}.} \label{NPPD}
\end{figure}

\begin{figure}
\centering
\includegraphics[width=40mm,height=40mm,angle=0]{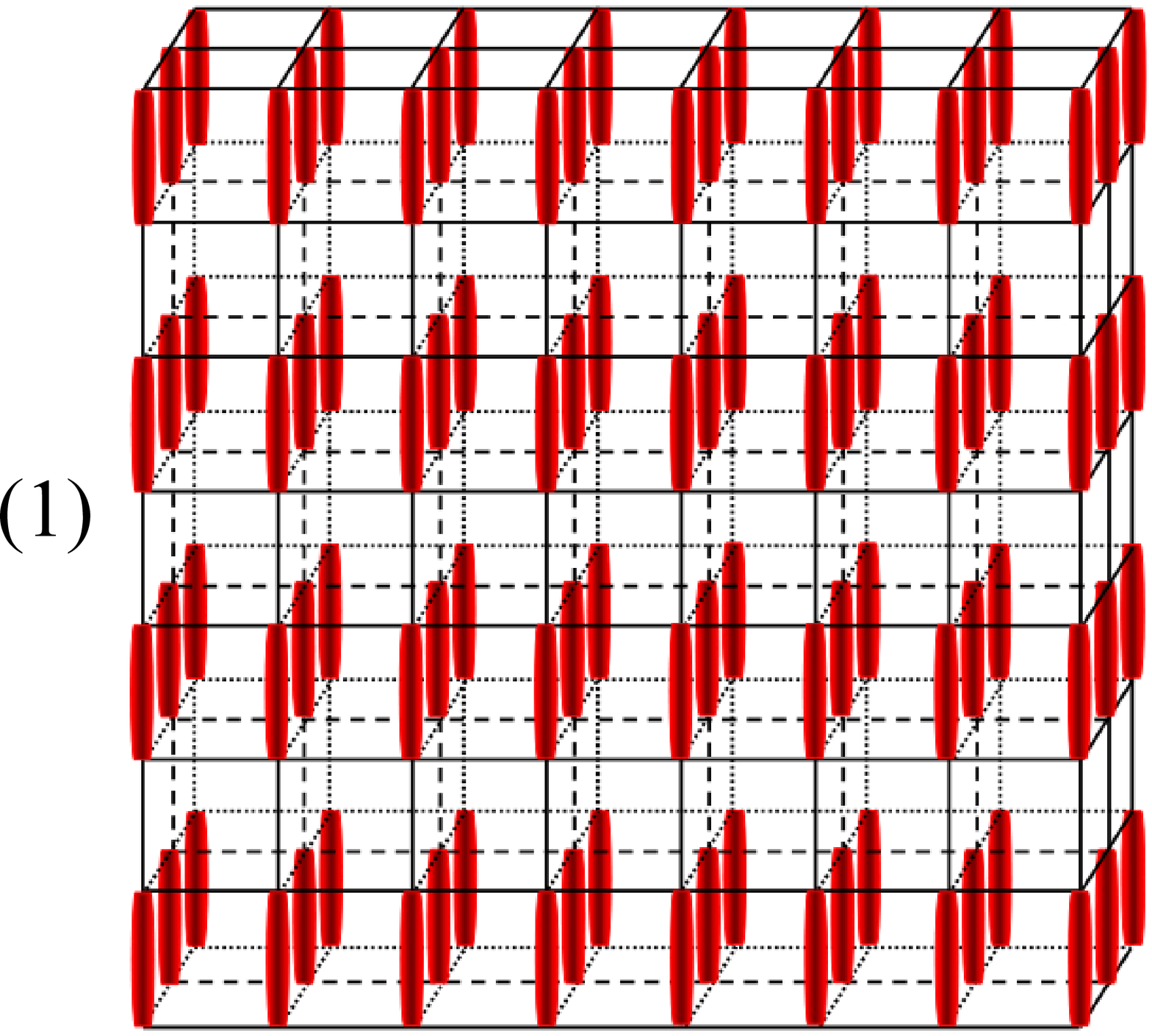}
\includegraphics[width=40mm,height=40mm,angle=0]{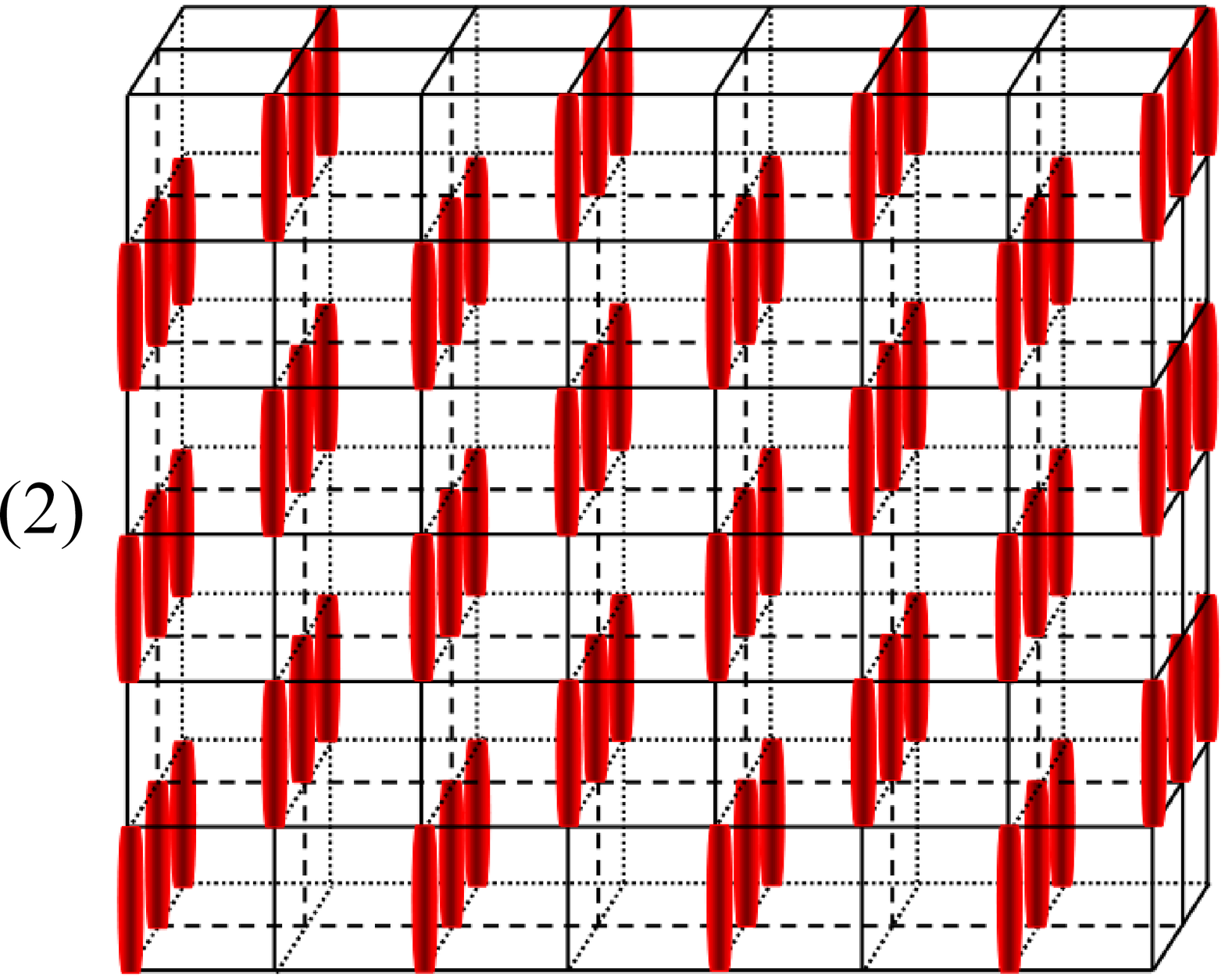}
\includegraphics[width=40mm,height=40mm,angle=0]{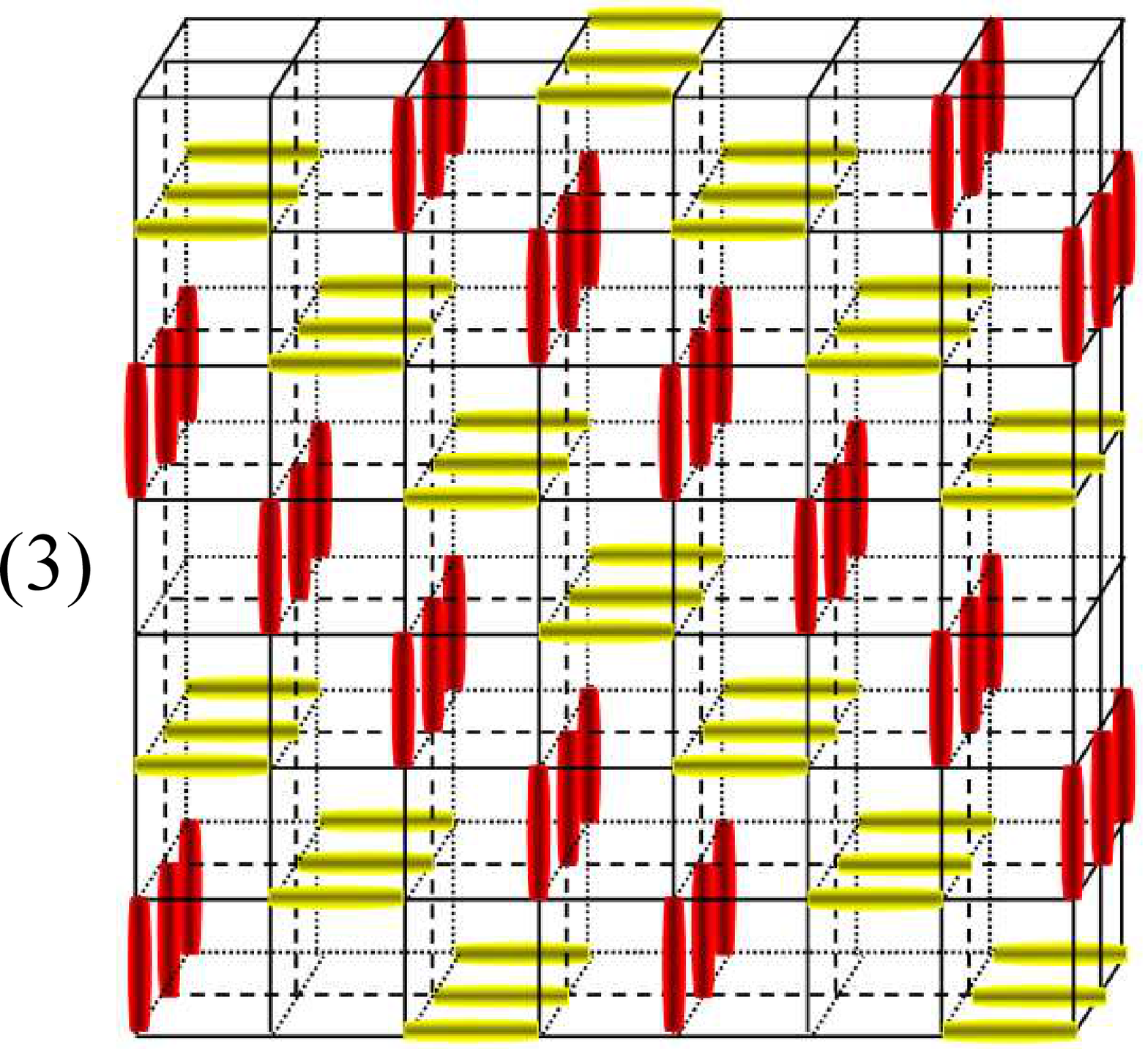}
\includegraphics[width=40mm,height=40mm,angle=0]{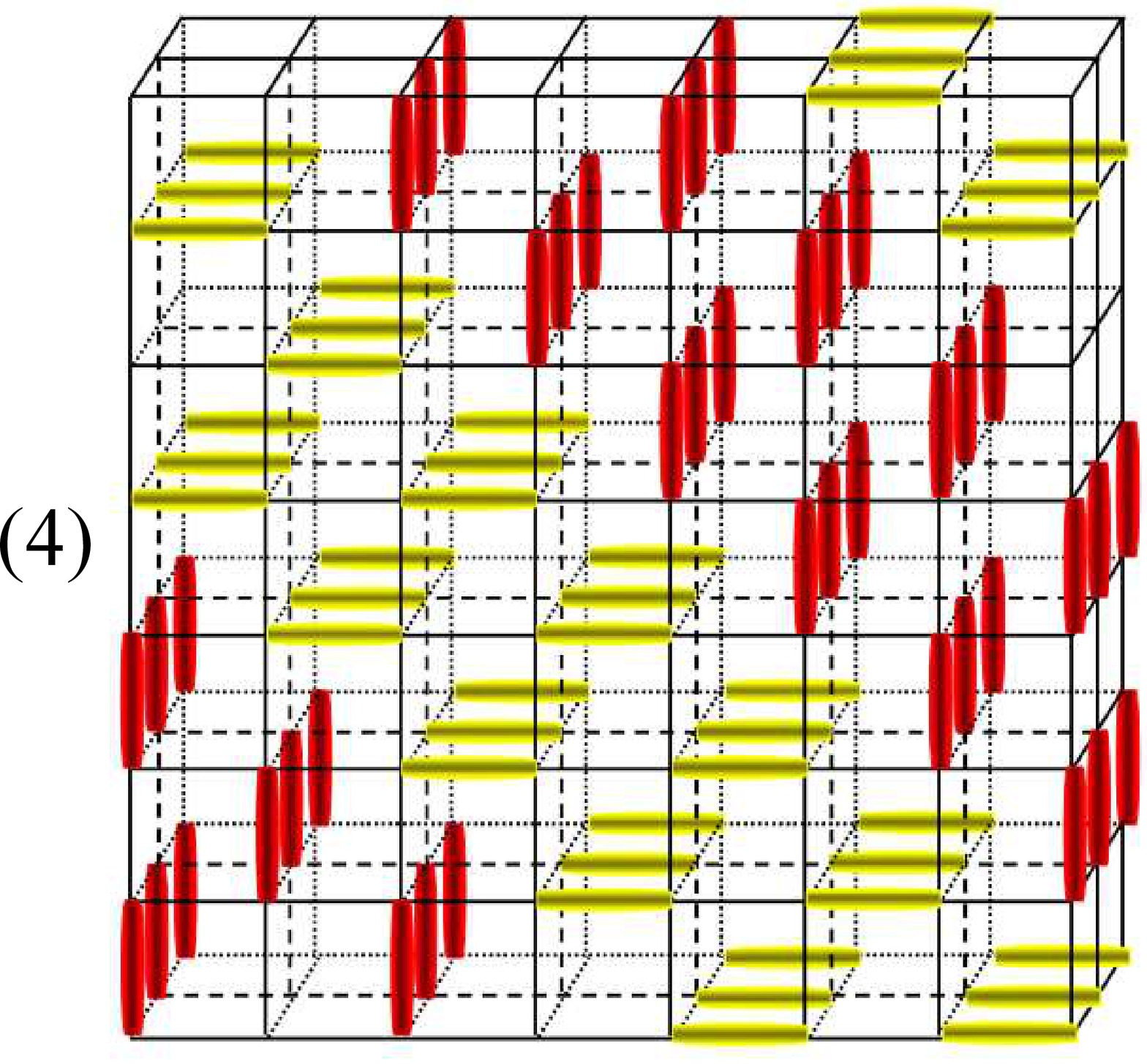}
\caption{(Color online) Non polarized dimers: Ground-state dimer configurations of the type 1, 2, 3 and 4 indicated in Fig. \ref{NPPD} are shown.} \label{NPGS}
\end{figure}

  For non zero $D$, the results obtained from the steepest descent method are shown in Fig. \ref{NPPD}.  The configurations 1 to 4 are shown in Fig. \ref{NPGS}. Note that for $D=0$, the GS is the uniform configuration of type 1,  similar to that in the polarized case.  We note the diagonal ordering of types 3 and 4.

 \subsection{Phase transition}
\begin{figure}
 \centering
 \includegraphics[width=60mm,angle=0]{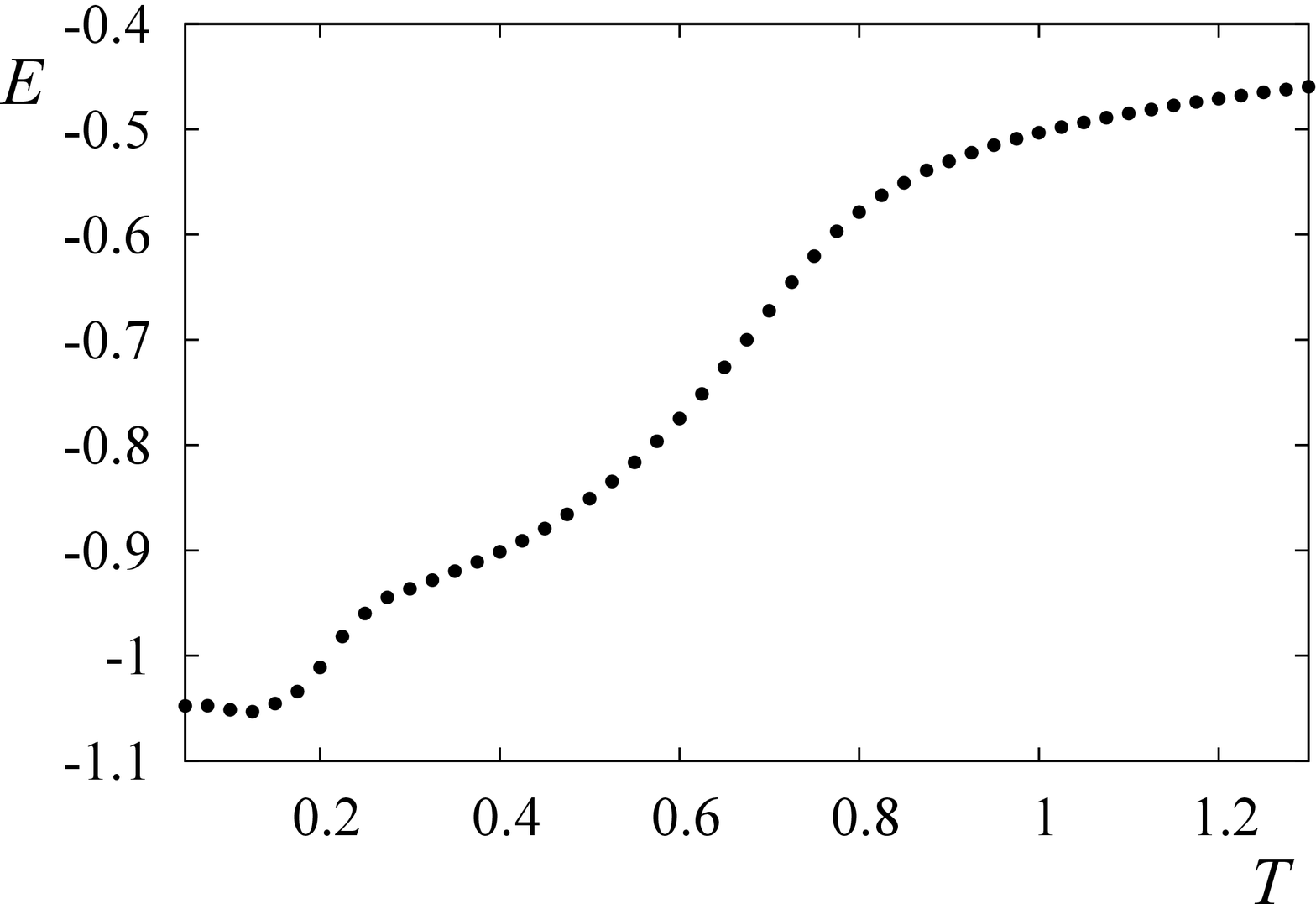}
  \includegraphics[width=60mm,angle=0]{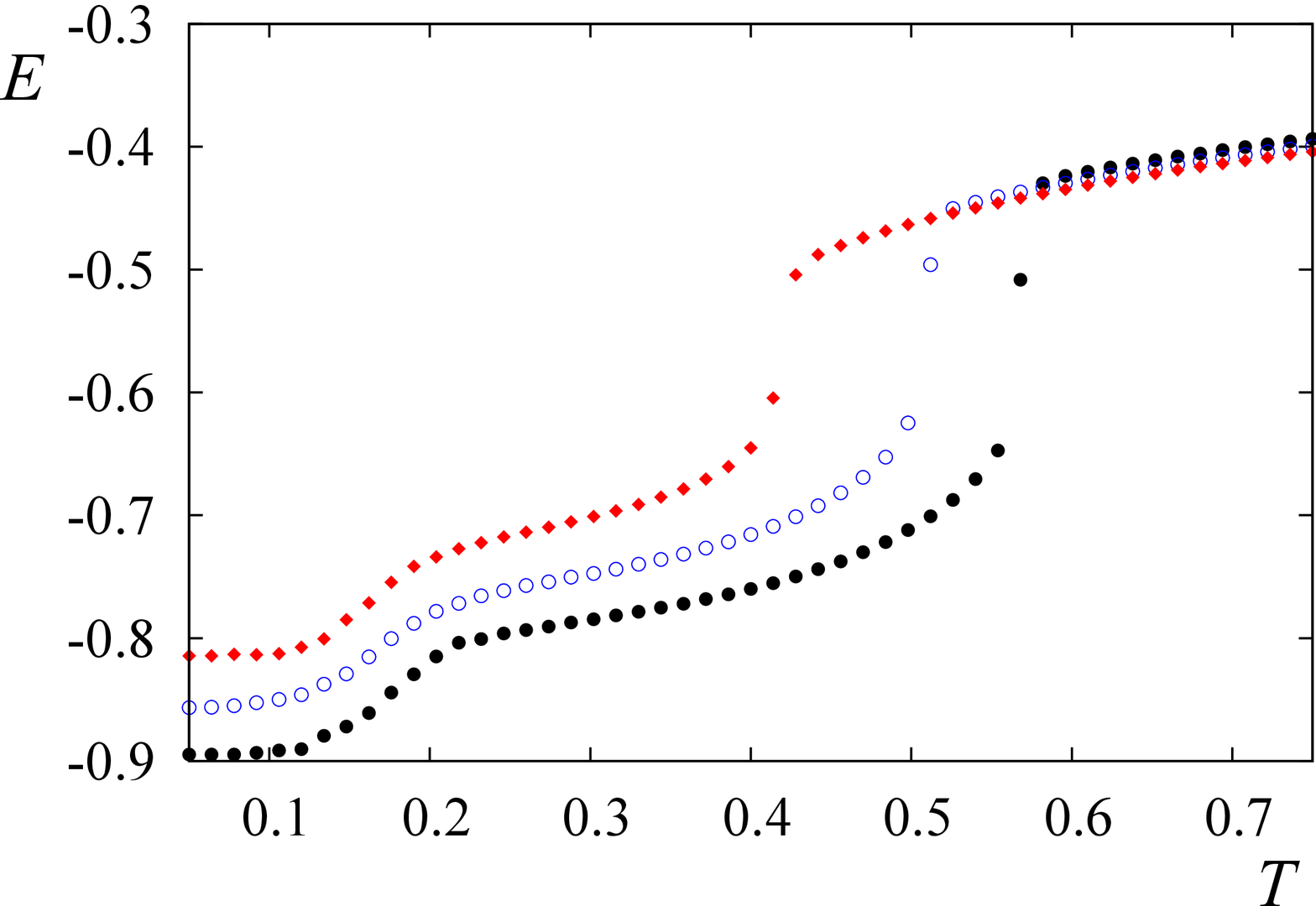}
 \caption{ (Color online) Non polarized dimers: Energy per dimer $E$  versus $T$ for (i) Top:  $D=0.4$  with  $r_c=2.5$ (top), (ii) Bottom: $D=0.6$  with  several $r_c$: 2.3 (black solid circles), 2.5 (blue void circles), 2.8 (red diamonds). } \label{NPD04}
\end{figure}

The energy versus $T$ is shown in Fig. \ref{NPD04} for two typical cases $D=0.4$ and $D=0.6$. Again here, we observe three successive phases with increasing $T$: for $D=0.4$ for example, the frozen phase is at $T<0.25$), ``compact" phase at ($0.25<T<0.7$) and isotropic phase at ($T>0.7$). For small $D$ (as shown with $D=0.4$), the transition is of second order, while for larger $D$ ($D=0.6$ for example), the transition becomes of first order. The energy histograms at the respective transition temperatures show a Gaussian peak for $D=0.4$ and a double-peak structure for $D=0.6$ (not shown).

As in the polarized case, the transition temperature decreases with decreasing $n_d$.
Note that the unfreezing is not a phase transition, it is a gradual process in which the maximum of the specific heat and the susceptibility as well as its unfreezing temperature do not depend on the lattice size, unlike in a true phase transition where these quantities depend on $L$.  As in the polarized case, the unfreezing is smoother in the case of smaller $n_d$ (not shown).


 \section{Concluding Remarks}\label{conclu}

 We have studied in this paper  a model of dimers moving on the simple cubic lattice. The dimers interact with each other via the nearest-neighbor 3-state Potts model of strength $J$ and competing interactions
 taken from a truncated dipolar interaction of amplitude $D$ and cutoff distance $r_c$.  The numerical steepest descent method has been used to determine the GS dimer configuration and MC simulations have been carried out to determine the ordered phase at finite $T$ and the nature of the phase transition.  We have found for both polarized and non polarized dimers complicated GS configurations as functions of ($D,r_c)$.  They are all compact, occupying a portion of the lattice.
 For small $D$,
 the GS is uniform,  and the surface separating the empty part of the lattice is flat.
 For large $D$, the GS is non uniform, the empty lattice space is a ``paraboloid hole".
 With increasing $T$, the dimers
 remain in  the  compact phase at low $T$. They undergo a
 transition to the spatially ``extended"
 isotropic phase at high $T$.  This transition has some resemblance with the unfolding transition of polymers
 in a solvent.     Note that the change from the compact phase to the isotropic phase is  of second order for small $D$ and of first order for large $D$.   The first-order transition observed here is due to the competition of the two terms in Eq. (\ref{dip}) for large enough $D$.
 We believe that other types of competing interactions which give rise to frustration and instability of the dimer configuration would cause a similar first-order ``unfolding" transition.

{}

\end{document}